\documentclass[letterpaper]{article}
\usepackage{xcolor}

\usepackage{flairs}

\usepackage{caption}
\usepackage{graphicx}
\usepackage{fancyhdr}
\usepackage{xcolor}
\usepackage{hyperref}
\usepackage{tabularx}
\usepackage{enumitem}
\usepackage{subcaption}
\usepackage{mdwmath}
\usepackage{amsmath,amssymb,amsfonts}
\usepackage{mathtools}
\usepackage{array}

\usepackage{times}
\usepackage{helvet}
\usepackage{courier}
\begin{document}
	\title{HARU-Net: Hybrid Attention Residual U-Net for Edge-Preserving Denoising in Cone-Beam Computed Tomography}
	\author{Khuram Naveed$^{1,*}$, and Ruben Pauwels$^1$\\
		\\
		$^1$Department of Dentistry and Oral Health, Aarhus University, Aarhus, 8000, Denmark\\
	}
	\maketitle
	\begin{abstract}
		\begin{quote}
			Cone-beam computed tomography (CBCT) is widely used in dental and maxillofacial imaging, but low-dose acquisition introduces strong, spatially varying noise that degrades soft-tissue visibility and obscures fine anatomical structures. Classical denoising methods struggle to suppress noise in CBCT while preserving edges. Although deep learning–based approaches offer high-fidelity restoration, their use in CBCT denoising is limited by the scarcity of high-resolution CBCT data for supervised training. To address this research gap, we propose a novel Hybrid Attention Residual U-Net (HARU-Net) for high-quality denoising of CBCT data, trained on a cadaver dataset of human hemimandibles acquired using a high-resolution protocol of the 3D Accuitomo 170 (J. Morita, Kyoto, Japan) CBCT system. The novel contribution of this approach is the integration of three complementary architectural components: (i) a hybrid attention transformer block (HAB) embedded within each skip connection to selectively emphasize salient anatomical features, (ii) a residual hybrid attention transformer group (RHAG) at the bottleneck to strengthen global contextual modeling and long-range feature interactions, and (iii) residual learning convolutional blocks to facilitate deeper, more stable feature extraction throughout the network. HARU-Net consistently outperforms state-of-the-art (SOTA) methods including SwinIR and Uformer, achieving the highest PSNR (37.52 dB), highest SSIM (0.9557), and lowest GMSD (0.1084). This effective and clinically reliable CBCT denoising is achieved at a computational cost significantly lower than that of the SOTA methods, offering a practical advancement toward improving diagnostic quality in low-dose CBCT imaging. 
		\end{quote}
	\end{abstract}

	\section{Introduction}
	Cone-beam computed tomography (CBCT) is among the most widely used imaging modalities in dentistry and otorhinolaryngology. It can provide three-dimensional visualization of the anatomical structures of the teeth and jaws, and ear-nose-throat (ENT) regions. CBCT enables volumetric assessment of anatomical structures, facilitating diagnosis and treatment planning across a wide range of applications, including endodontics, implantology, orthodontics, temporomandibular joint (TMJ) evaluation, maxillofacial surgery, and ENT imaging. These applications are made possible by relatively low radiation doses in CBCT imaging and high spatial resolution, which make it particularly suitable for imaging the head and neck regions.
	
	Despite its broad clinical utility, the requirement to maintain low radiation doses to minimize patient harm results in a relatively high degree of noise in CBCT scans \cite{scarfe2008cone,pauwels2015technical,pauwels2012effective,schulze2011artefacts}. These noise patterns are composed of quantum noise due to low exposure \cite{bryce2021low,goldman2007principles,pauwels2016reduction}, and additive noise due to sensor electronics and post-processing errors (e.g., quantization errors) \cite{nuyts2013modelling,kalender2011computed}. Such granular noise often compromises the visual clarity of anatomical structures, causing soft-tissue regions to become largely indistinguishable and impairing the visualization of hard-tissue boundaries and small lesions \cite{mahesh2022prevalence,olszewski2020artifacts}. Such limitations in CBCT imaging reduce diagnostic confidence and often necessitate repeated scans or supplementary imaging with CT or MRI. These challenges underscore the need for efficient post-processing noise reduction techniques that can enhance image quality without increasing the radiation dose. As the clinical use of CBCT increases, the demand for reliable, high-quality CBCT denoising solutions has grown substantially.
	
	Since acquisition-related noise is not unique to CBCT imaging, denoising has long been employed to address the low signal-to-noise ratio (SNR) problem (e.g., \cite{naveed2025naada,khawaja2019improved,naveed2021towards,naveed2019multiscale}). In this context, deep learning (DL) approaches for CBCT image enhancement have been explored in some recent studies, e.g., \cite{yunker2024Unet3D4cbctDen1,zhao2025Unet3D4cbctDen2,naveed2023Unet}. However, the literature remains limited when considering CBCT used in dentistry and ENT imaging, with the majority of studies focused on CBCT used in radiotherapy. This scarcity stems from the dependence of supervised learning on paired high-quality reference data, which requires acquiring high-dose scans. Such low- and high-radiation image pairs are ethically and clinically impractical. Notably, even the supervised approaches reported in \cite{yunker2024Unet3D4cbctDen1,zhao2025Unet3D4cbctDen2} were only made possible through a dataset temporarily released as part of the Grand Challenge in \cite{ICASSP_Grand_Challenge}. Although phantom or cadaver datasets with controlled exposure levels can, in principle, support supervised training, they fail to fully capture the anatomical variations and tissue heterogeneity present in real patients. A way around this is to use self-supervised learning frameworks, as demonstrated in \cite{choi2021selfSupCBCTDEn,yun2023cbctnoise2void,zanini2024enhancingCBCT}. These frameworks train neural networks to map noisy inputs to independently corrupted versions of the same data, which facilitates the noise being averaged out without the need for clean supervision. However, the performance of self-supervised learning methods tends to be suboptimal.
	
	This work seeks to address this challenge through a cadaver-based CBCT dataset acquired under high-dose settings, with simulated noise added to generate paired noisy and clean CBCT slices. We introduce a pre-processing pipeline based on morphological operations for detection and segmentation of tissue from background (air), to ensure that the latter is ignored for downstream tasks. For edge-preserving denoising, we propose a hybrid attention residual U-Net (HARU-Net), which is capable of capturing anatomical structures and subtle tissue variations across CBCT slices, thereby enabling robust recovery of structural details within complex noise patterns. For validation and efficiency, we compare our method against state-of-the-art (SOTA) transformer-based architectures for denoising, i.e., SwinIR \cite{liang2021swinir} and Uformer \cite{wang2022uformer}, and a residual U-Net as a standard CNN benchmark method.
	
	\section{Methods}
	This section describes: (i) the dataset and pre-processing steps for segmentation of the foreground tissue, (ii) preparation of training data through dynamic patching and the addition of noise, and (iii) a detailed explanation of the proposed HARU-Net architecture.
	
	\subsection{Pre-processing and Data Preparation}
	Here, we describe the preparation of noisy and clean image pairs and dynamic patching developed using only the segmented anatomical regions.
	
	\subsubsection{Dataset Description}
	The dataset employed in this study consists of CBCT scans of human hemimandibles, which were acquired at Chulalongkorn University using the 3D Accuitomo 170 CBCT (J. Morita, Kyoto, Japan) system. A total of $21$ hemimandibular specimens were obtained from the Department of Anatomy, Chulalongkorn University. The study protocol received a certificate of exemption from the Faculty of Dentistry, Chulalongkorn University Human Research Ethics Committee (approval code HREC-DCU 2015-032). The imaging protocol followed standard adult high-resolution exposure settings: $90$ kV, $5$ mA, and an exposure time of $30.8$ seconds per scan. The scans were reconstructed using a field of view (FOV) of $5 \times 5$ cm at an isotropic voxel resolution of $0.08$ mm, allowing for high-detail visualization of osseous structures.
	
	Each of the $21$ CBCT volumes was sliced along the frontal, axial, and sagittal planes using ImageJ (National Institutes of Health, Bethesda, MD, USA) to obtain 2D views from three anatomical directions. This resulted in a total of $26{,}317$ 2D slices used for training and testing. Subsequently, each of these slices was pre-processed to extract the relevant anatomical regions for model input preparation, as detailed in the next section.
	
	\subsubsection{Preparation of Noisy Samples}
	Noise in reconstructed CBCT images is primarily composed of two sources: (i) the stochastic nature of discrete photon detection (quantum noise) and (ii) electronic circuitry used for photon detection (electronic noise).
	
	Quantum noise depends on the radiation dose; that is, a higher radiation dose ensures a higher SNR due to a larger number of photon detections, with SNR being proportional to the square root of the number of photons (and thus, the dose). While the number of detected photons $N_P$ follows a Poisson distribution, $N_P \sim \mathrm{P}(\lambda)$, where the mean photon count $\lambda$ is proportional to the local X-ray intensity incident on the detector, for large photon flux, the Poisson process can be approximated by a Gaussian distribution with variance equal to its mean, $\mathrm{Var}(N) = \lambda$. However, after logarithmic transformation and filtered back-projection, this variance is propagated and reflected in the reconstructed voxel intensities. To approximate this behavior at the image level, the quantum noise can be expressed as an additive Gaussian term with uniform variance $\mathcal{N}(0, \sigma_q^2)$, where $\sigma_q$ controls the strength of the quantum noise contribution in the reconstructed domain, which depends on the radiation dose \cite{pauwels2015technical,kak2001principles}.
	
	Fluctuations in the detection circuitry, thermal fluctuations, and analog-to-digital conversion contribute to electronic noise, which is independent of the signal and is commonly approximated as a zero-mean Gaussian process $\mathcal{N}(0, \sigma_e^2)$, with $\sigma_e$ denoting the standard deviation corresponding to the level of electronic noise.
	
	The recorded voxel values in the reconstructed CBCT volume $I'$ can be modeled as the sum of the true values $I$ and these two additive noise terms, given as follows:
	\begin{equation}
		I' = I + \psi_q + \psi_e,
	\end{equation}
	where $\psi_q \sim \mathcal{N}(0, \sigma_q^2)$ denotes the quantum noise component and $\psi_e \sim \mathcal{N}(0, \sigma_e^2)$ denotes the electronic noise component in a voxel.
	
	\subsubsection{Segmentation of Anatomical Regions and Dynamic Patching}
	This section describes an unsupervised clustering and morphological operations pipeline designed specifically to isolate the foreground anatomical regions from background air in CBCT slices. This approach facilitates dynamic placement of patches exclusively within anatomically meaningful areas, thereby effectively excluding empty, irrelevant air regions.
		\begin{figure*}
		\centering
		\centering
		\includegraphics[width=\textwidth]{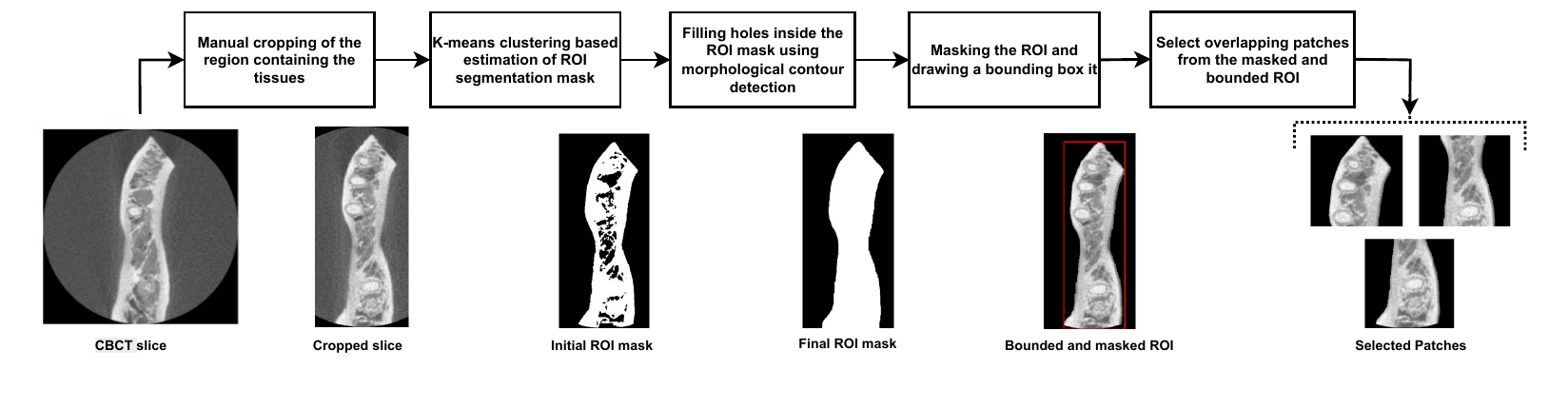}
		\caption{An illustration of the proposed pre-processing pipeline for detection and segmentation of the foreground-tissue from the background air and noise.}
		\label{fig01}
	\end{figure*}

	\begin{enumerate}
		\item \textit{Manual Cropping for Excluding Empty Regions}  
		The first step of this pipeline was performed manually, where all slices were cropped using ImageJ software. The cropping area was determined by visually inspecting all slices to identify the region encompassing the highest anatomical interest. This ensured that each cropped slice contained the relevant anatomical structures while maintaining uniform dimensions across the dataset.
		
		\item \textit{K-Means Clustering for Foreground Segmentation}  
		We employed an unsupervised K-means clustering algorithm \cite{ajala2012fuzzy} to isolate foreground tissue pixels, leveraging their significantly higher intensity values compared to the relatively low pixel values in air- and noise-dominated regions. The algorithm was applied with $k = 2$ and a convergence criterion combining a maximum of 100 iterations and a minimum cluster center shift of $0.2$ (i.e., $\epsilon = 0.2$). Given that $I_k \in \mathbf{I}$ denotes the $k^{\text{th}}$ slice in the 3D CBCT volume $\mathbf{I}$, the algorithm extracts non-zero pixel intensities and reshapes them into a one-dimensional vector $\Tilde{I}_k$.
		
		The final output is an initial binary mask $M_0 \in \{0, 1\}^{H \times W}$, where the foreground pixels represent the anatomical regions of interest. However, the resulting mask may contain distorted boundaries between air and tissue due to partial volume averaging, as well as voids within the anatomical structures due to pockets filled with air, low-density tissues, or noise. The following two steps address this issue.
		
		\item \textit{Morphological Dilation}  
		We employ a morphological dilation operation to correct the distorted boundaries as well as internal noise pockets of the binary mask $M_0$. The dilation operation is applied using a square structuring element (of size $5 \times 5$) to smooth boundary irregularities by expanding the segmented regions. This operation is also helpful in filling narrow gaps, resulting in a more contiguous representation of the anatomical structures in the improved mask, denoted by $M_1$, along with regularized tissue boundaries. However, this step does not fully address the discontinuities caused by larger empty pockets within the anatomical boundaries.
		
		\item \textit{Hole Detection and Filling via Contour Hierarchy}  
		This step addresses voids or holes too large to fill using dilation. We detect these empty regions using a hierarchical contour retrieval method ((\verb|cv2.RETR_CCOMP|)), which captures both outer and inner contours \cite{suzuki1985topological}. Inner contours denote the hollow or empty regions, while their surrounding regions are termed the parent contours. Once all inner contours are identified, we fill them using a region-growing method, namely the flood-fill approach \cite{haralick1985image}. This is further supplemented by morphological smoothing operations to obtain a smooth mask and close residual gaps. The process iteratively increases the kernel size, starting from $15 \times 15$ with a step size of 5, until either no new contours are detected or a maximum iteration count is reached. This iterative refinement produces a final binary mask $M_f$, where anatomical regions are fully enclosed and holes are eliminated.
		
		\item \textit{Bounding Box Extraction and Dynamic Patching}  
		To ensure that patches used for training the DL models are selected exclusively from the foreground tissue regions, the contours of these regions are extracted to define axis-aligned bounding boxes around each fully connected component. These bounding boxes localize the segmented anatomical areas, thereby enabling dynamic patch placement within the tissue regions and preventing the inclusion of predominantly empty areas.
	\end{enumerate}
	
	However, this approach may introduce redundancy when smaller anatomical regions are completely enclosed within the bounding box of a larger region. To mitigate this, nested bounding boxes (i.e., boxes entirely contained within another) are identified and removed by evaluating the spatial containment relationships among all boxes. Consequently, only the largest non-overlapping bounding boxes are retained for patch extraction.
	
	Subsequently, non-overlapping patches of size $256 \times 256$ pixels are selected starting from the top-left corner of each slice. In cases where a few rows or columns remain uncovered after patch selection, additional overlapping patches are generated to ensure complete coverage of the anatomical area. In cases where a bounding box has dimensions smaller than the desired patch size, it is symmetrically extended to meet the minimum size requirement while ensuring it remains within the image boundaries.
	
	A total of $50{,}026$ pairs of noisy and clean patches were generated following the process detailed above on $14$ randomly selected CBCT volumes (i.e., $\approx 70\%$ of the total data). For validation and testing, a total of $8{,}971$ and $10{,}462$ patches were obtained from $3$ randomly selected CBCT volumes each ($\approx 15\%$ of the total data).
	
	\begin{figure*}
		\centering
		\centering
		\includegraphics[width=\textwidth]{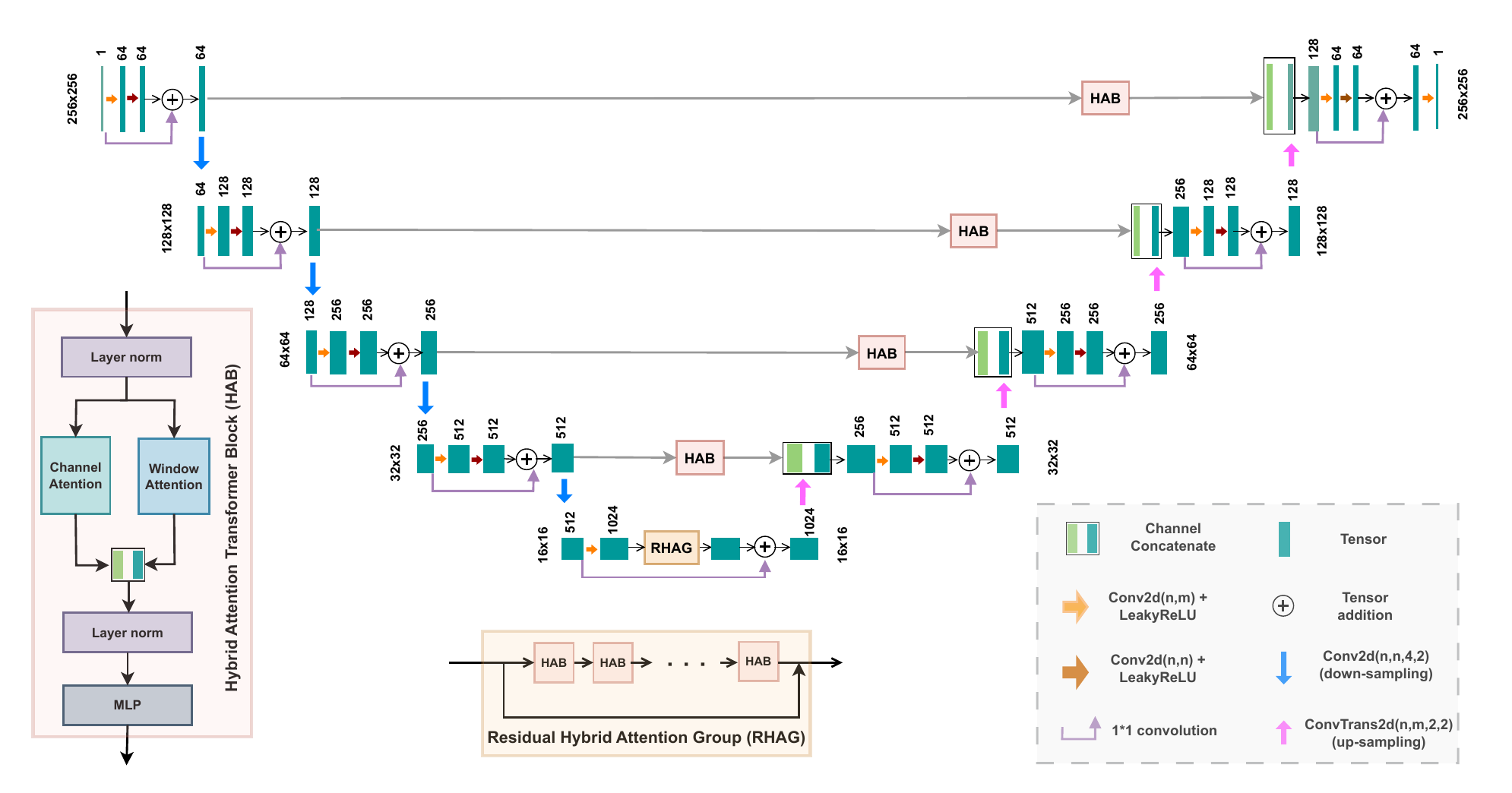}
		\caption{Architecture of the proposed HARU-Net, incorporating hybrid attention transformer modules (HABs and RHAGs) within the skip connections and bottleneck of a residual U-Net to improve feature representation.}
		\label{fig02}
	\end{figure*}
	
	\subsection{Hybrid Attention Residual U-Net}
	This section introduces the proposed hybrid attention residual U-Net (HARU-Net), illustrated in Fig.~\ref{fig02}, which integrates transformer-based attention blocks within the classical encoder–decoder framework of U-Nets to enhance representational capacity for the challenging task of image denoising. Conventional convolutional blocks primarily rely on localized filtering operations that act as low-frequency estimators and, by design, tend to suppress high-frequency details such as edges and corners. However, preservation of these high-frequency components is critical for effective denoising, as they encode essential structural and anatomical information. To address this limitation, attention mechanisms and transformer architectures have been increasingly adopted to model global dependencies and retain high-frequency details. In this context, HARU-Net incorporates a hybrid attention block (HAB) and a residual hybrid attention group (RHAG), originally introduced in a recent transformer-based architecture for image super-resolution and denoising~\cite{chen2023hat}, to activate a larger number of pixels and enable effective modeling of global dependencies. Such collaborative use of convolutional and transformer blocks helps achieve effective denoising while maintaining computational efficiency compared with fully transformer-based solutions.
	
	The proposed HARU-Net architecture is composed of four main types of convolutional and transformer-based blocks: (i) residual convolutional encoding blocks for robust feature extraction, (ii) hybrid attention transformer blocks (HABs) integrated into the skip connections to emphasize relevant features at each resolution level, (iii) a residual hybrid attention group (RHAG) at the bottleneck to enhance the representational capacity of the deepest feature maps, and (iv) residual convolutional decoding blocks that progressively reverse the encoding process and reconstruct high-resolution feature maps with the assistance of the HABs and RHAG. This architecture effectively combines the strengths of convolutional layers for local feature estimation with the enhanced representational capacity of transformer-based attention modules for capturing contextual dependencies across scales. As a result, HARU-Net is able to suppress background noise while preserving fine anatomical details critical for CBCT interpretation. The following subsections describe each architectural component in detail.
	
	\subsubsection{Encoder}
	The encoder consists of four stages, each containing a residual convolutional encoding block composed of two convolutional layers in series, each formed by two $3\times3$ convolutions followed by a LeakyReLU activation, and a $1 \times 1$ convolutional projection of the input to the output serving as a skip connection to feed forward dimension-aligned features for residual learning. This residual design ensures stable gradient flow and effective feature propagation, even when the number of channels changes. At each stage, the first convolution operation doubles the number of channels (from $c$ to $2c$, $4c$, and $8c$), where $c$ denotes the number of channels after the first convolution operation in the initial block. After each residual block, the spatial resolution of the output tensor is reduced by a factor of $2$ using learnable $4\times4$ convolutions with a stride of $2$. We used LeakyReLU activation due to its sensitivity to negative feature values, enabling enhanced feature flow into the deeper layers. Furthermore, the use of convolutional operations for down-sampling, instead of pooling operations, preserves more signal information and enables flexible feature compression by learning data-driven down-sampling filters.
	
	\subsubsection{Hybrid Attention Transformer Block}
	The Hybrid Attention Block (HAB), introduced in \cite{chen2023hat}, combines window-based self-attention with channel attention to jointly capture both local and global contextual information. The windowed self-attention mechanism, originally introduced in the Swin Transformer architecture \cite{liu2021swin}, adapts standard self-attention to a local scale through window partitioning while retaining longer-range contextual dependencies via overlapping and shifted windows across layers. The hybrid attention module reinforces the global feature representation by incorporating a channel attention component that re-weights feature channels according to their global relevance, thereby strengthening the model’s ability to emphasize contextually important information and activate a larger proportion of input pixels. We present a brief description of windowed and channel attention for necessary background as follows:
	
	\begin{enumerate}
		\item \textit{Windowed self-attention} \cite{liu2021swin} is a localized variant of the standard self-attention mechanism \cite{vaswani2017attention}, designed to capture fine-grained spatial patterns in visual data. Serving as a core building block of the Swin Transformer architecture \cite{liu2021swin}, this approach partitions the input image into fixed-size windows and computes self-attention independently within each window, enabling efficient modeling of local context while progressively aggregating global information across layers. Within each window, feature embeddings are projected into query, key, and value representations, and attention weights are computed based on pairwise query–key similarity to generate context-aware features. By operating on localized neighborhoods, windowed self-attention preserves fine structural and textural details useful for image restoration. Moreover, the use of shifted windows across consecutive layers facilitates information exchange between neighboring windows, effectively integrating local details with global anatomical context.
		
		\item \textit{Squeeze-and-excitation channel attention} enhances feature representation by adaptively re-weighting the importance of each feature channel. It models inter-channel dependencies by aggregating global spatial information through operations such as global average pooling, followed by a small multi-layer perceptron (MLP) that learns attention weights. These learned weights are then used to selectively emphasize informative channels and suppress less relevant ones, thereby improving the network’s ability to focus on diagnostically significant features while reducing redundancy.
	\end{enumerate}
	
	A sketch of the internal structure of the HAB is illustrated in Fig.~2 (left), where the hybrid-attention mechanism is positioned between two consecutive layer normalization (LN) operations and followed by a multilayer perceptron (MLP) with nonlinear activation. This design ensures stable feature transformations and enhanced learning of pixel-wise activations. In this way, the efficacy of a recent high-fidelity transformer block is embedded within a robust residual convolutional architecture, enabling enhanced activation of features at both local and global scales and effectively addressing the limitations imposed by localized receptive fields.
	
	We employ the HABs directly at the skip connections to refine and enhance the core feature representations transmitted between the encoder and decoder stages. Readers interested in learning more about HAB and its ability to enhance receptive fields can refer to \cite{chen2023hat}.
	\begin{table*}[h!]
		\centering
		\caption{Evaluation of denoising performance on the test dataset.}
		\begin{tabular}{|c|c|c|c|c|c|c|}
			\hline
			\textbf{Metric}  & \textbf{ResUNet} & \textbf{Uformer} & \textbf{SwinIR} & \textbf{HAT} & \textbf{HARU-Net} \\
			\hline
			\textbf{PSNR}  & 35.03 & 36.25 & 36.12 & 36.70 & \textbf{37.52} \\
			\hline
			\textbf{SSIM}  & 0.9542 & 0.9447 & 0.9551 & \textbf{0.9569} & 0.9557 \\
			\hline
			\textbf{GMSD}  & 0.1240 & 0.1147 & 0.1151 & 0.1119 & \textbf{0.1084} \\
			\hline
		\end{tabular}
		\label{table1}
	\end{table*}
	\begin{table*}[h!]
		\centering
		\caption{Computational costs during inference}
		\begin{tabular}{|c|c|c|c|c|c|}
			\hline
			
			\textbf{Measure} & ResUnet & Uformer & SwinIR & HAT& HARU-Net \\ \hline
			
			\textbf{Flops per patch (GMACs)} & \textbf{6.898} & 78.027 & 111.069 & 349.358 & 40.760\\ 
			\hline
			\textbf{Time per scan (minute)} & \textbf{0.205} & 4.298 & 8.852 & 13.095 & 1.985\\
			\hline
			
		\end{tabular}
		\label{table2}
	\end{table*}
	\subsubsection{RHAG and Bottleneck}
	We employ a more extensive and robust attention-based feature modeling framework, namely the residual hybrid attention transformer group (RHAG), at the bottleneck, where the feature representations are at their deepest level. This design enables the modeling of long-range contextual dependencies at a relatively low additional computational cost. It is important to note that the proposed RHAG is a simplified variant of the original RHAG module introduced in \cite{chen2023hat}, where only the HAB block is repeated six times in series within a residual configuration, allowing the network to capture higher-order global patterns while preserving data fidelity through residual learning.
	
	Before RHAG operation, the bottleneck employs a convolutional operation to increase the number of channels from $512$ to $1024$ for the resulting feature map. Similar to the encoding blocks, the bottleneck also incorporates a residual connection to facilitate effective gradient propagation and stable training.
	
	\subsubsection{Decoder with Skip Connections}
	The decoder mirrors the encoder, progressively restoring spatial resolution through transposed convolutions followed by convolutional refinement. To enhance information flow through the skip connections, an HAB is integrated at each skip pathway to refine the transmitted features. In doing so, the encoder features are selectively enhanced, emphasizing salient anatomical structures over noise and thereby improving the reconstruction of fine details. After concatenation, the fused feature maps are further refined using a residual convolutional block before being forwarded to the next decoding stage.
	
	\subsection{Training and hyper-parameters}
	Mean Squared Error (MSE) was used as the loss function to train the proposed HARU-Net. The network was trained using the Adam optimizer with an initial learning rate of $1\times10^{-4}$. A learning rate scheduler reduced the learning rate when the validation loss failed to improve for five consecutive epochs, helping the model escape shallow minima and stabilize convergence. Training was terminated early if no improvement in validation loss was observed for $20$ epochs, preventing overfitting and unnecessary computation. The same optimizer, learning-rate scheduling, and early-stopping strategy were consistently applied across all comparative methods (described in the next section) to ensure a fair evaluation.

	\begin{figure*}
		\centering
		\begin{subfigure}[b]{0.4\textwidth}
			\includegraphics[width=\textwidth]{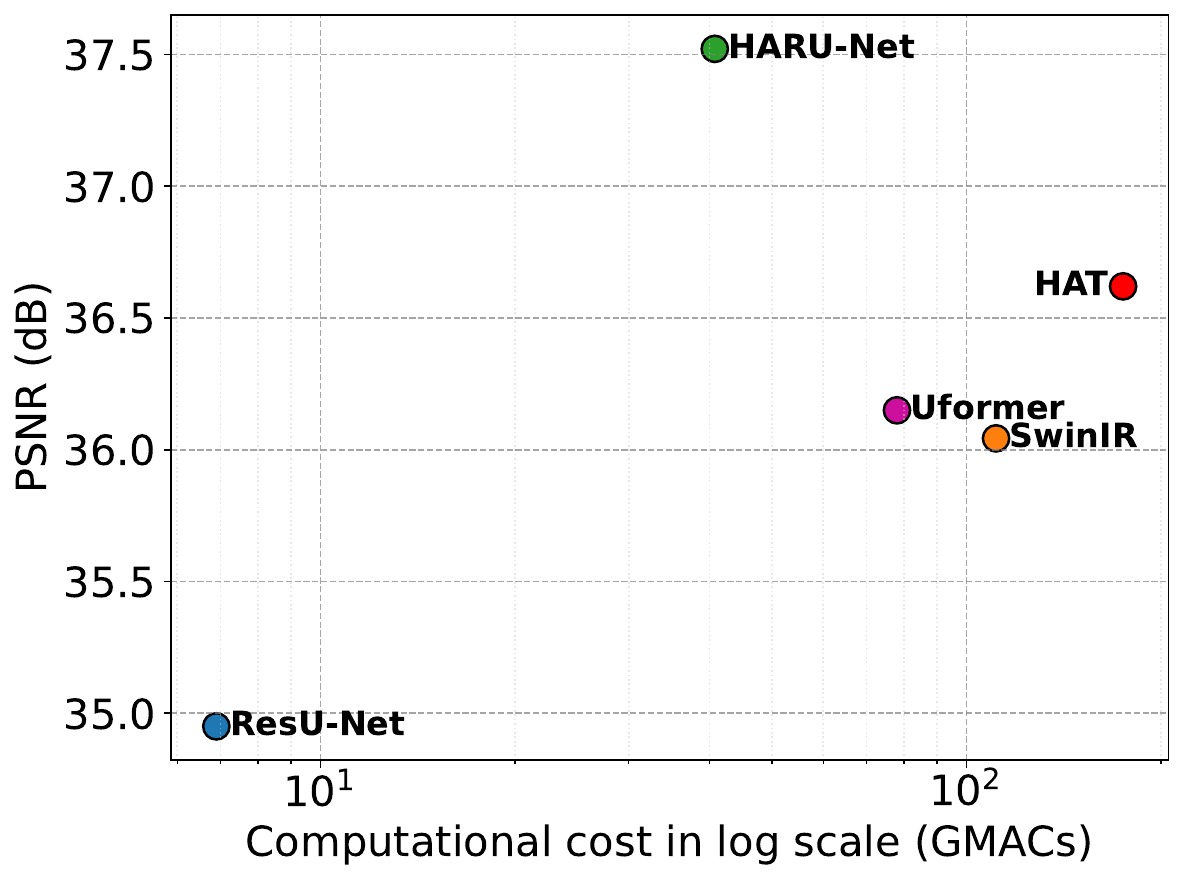}
			\label{figVR1a}
		\end{subfigure}
		\begin{subfigure}[b]{0.4\textwidth}
			\includegraphics[width=\textwidth]{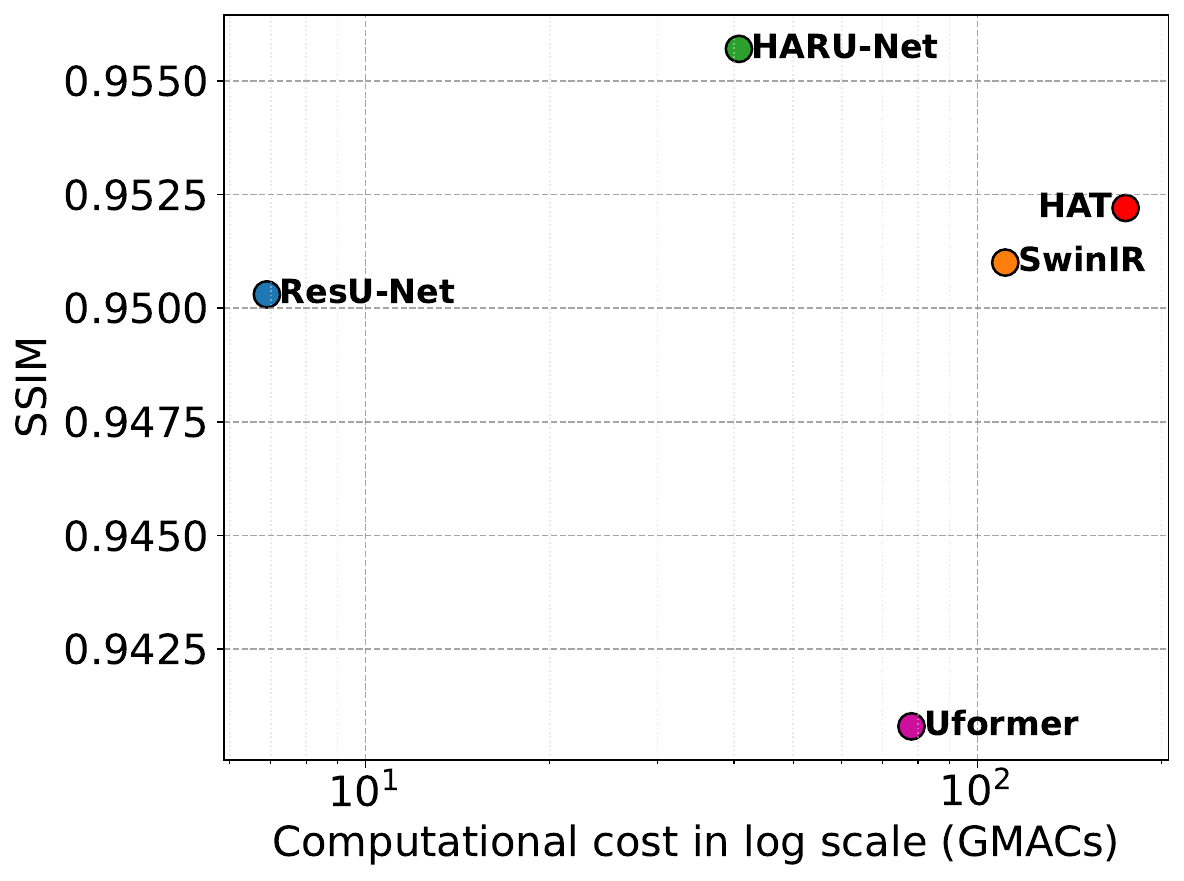}
			\label{fig5b}
		\end{subfigure}
		\caption{Comparison of computational cost with the performance in terms of PSNRs and SSIMs.}
		\label{fig03}
	\end{figure*}
	
	\subsection{Experimental setup}
	This section presents the experimental setup and materials used to evaluate the performance of the proposed HARU-Net against several state-of-the-art (SOTA) denoising methods. To ensure a comprehensive comparison, we selected algorithms that are widely recognized for their denoising effectiveness across medical, dental, and natural image domains. The comparison pool includes SwinIR \cite{liang2021swinir}, a Swin Transformer–based image restoration model known for its strong performance in diverse imaging applications. We additionally employ Uformer \cite{wang2022uformer}, a U-shaped transformer architecture inspired by the U-Net design but enhanced with transformer-based feature modeling. To provide an architectural baseline, we also include a Residual U-Net (ResU-Net), which forms the foundational backbone of HARU-Net shown in Fig.~\ref{fig02} with the HAB and RHAG blocks removed.
	
	For quantitative evaluation of the denoising performance of the comparative methods, we employ three widely used image quality metrics and a fourth metric to compare computational cost:
	\begin{enumerate}
		\item \textit{Peak Signal-to-Noise Ratio} (PSNR) quantifies the pixel-wise agreement between the noise-corrected image and the reference image (i.e., the original CBCT data before noise addition). It is derived from the MSE and is expressed in decibels (dB). Higher PSNR values indicate closer correspondence to the reference image, reflecting effective noise suppression. PSNR is defined as
		\begin{equation}
			\mathrm{PSNR} = 10 \log_{10}\left( \frac{L^2}{\mathrm{MSE}} \right),
		\end{equation}
		where $L$ denotes the maximum possible pixel intensity value of the image (e.g., $L = 255$ for an image represented in $8$-bit unsigned integer format, while $L = 1$ for a normalized image), and $\mathrm{MSE} = \frac{1}{N} \sum_{i=1}^{N} \left( I(i) - \hat{I}(i) \right)^2$, with $I$ and $\hat{I}$ respectively denoting the reference and denoised images, each having $N$ pixels.
		\item \textit{Structural Similarity Index Measure} (SSIM) evaluates perceptual similarity by jointly assessing luminance, contrast, and structural consistency between two images. SSIM is more closely aligned with human perception and understanding than PSNR. Since anatomical structures and contrast preservation are critical in CBCT, SSIM is a vital indicator of preserved structural information. Its values range between $0$ and $1$, with values close to $1$ indicating better structural preservation. SSIM can be defined as 
		\begin{equation}
			\mathrm{SSIM}(I,\hat{I}) = 
			\frac{(2\mu_{I} \mu_{\hat{I}} + C_1)(2\sigma_{I,\hat{I}} + C_2)}
			{(\mu_{I}^2 + \mu_{\hat{I}}^2 + C_1)(\sigma_{I}^2 + \sigma_{\hat{I}}^2 + C_2)},
		\end{equation}
		where $\mu_{I}$ and $\mu_{\hat{I}}$ denote the mean intensities, $\sigma_{I}^2$ and $\sigma_{\hat{I}}^2$ denote the variances, and $\sigma_{I,\hat{I}}$ denotes the covariance of the reference and denoised images $I$ and $\hat{I}$. The constants $C_1$ and $C_2$ are included to stabilize the division.
		\item \textit{Gradient Magnitude Similarity Deviation} (GMSD) measures the deviation between the gradient magnitudes of the reference and the noise-corrected image. GMSD is particularly useful for assessing edge preservation and retention of fine structural details, which are essential for accurate interpretation of CBCT scans. Lower GMSD values indicate better preservation of anatomical boundaries and textures.
		
		Mathematically, GMSD of the reference and denoised images $I$ and $\hat{I}$ is given as follows:
		\begin{equation}
			GMSD = \sqrt{ \frac{1}{N} \sum_{i=1}^{N} \left( \nabla_{I\sim\hat{I}}(i) - \overline{\nabla}_{I\sim\hat{I}} \right)^2 },
		\end{equation}
		where $\nabla_{I\sim\hat{I}}(i)$ denotes the gradient magnitude similarity at the $i$th pixel in images $I$ and $\hat{I}$, defined as
		\begin{equation}
			\nabla_{I\sim\hat{I}}(i) = \frac{2 \nabla_{I}(i) \nabla_{\hat{I}}(i) + C}
			{\nabla_{I}(i)^2 + \nabla_{\hat{I}}(i)^2 + C},
		\end{equation}
		where $\nabla_{I}$ and $\nabla_{\hat{I}}$ denote the gradients of the reference and denoised images $I$ and $\hat{I}$, respectively, and $C$ is a small positive constant.
		\item \textit{Giga-Multiply-Accumulate operations} (GMACs) quantify the computational cost (i.e., one GMAC is equal to two giga floating point operations per second (FLOPS)) required for a single forward pass through the model. This metric reflects the inference efficiency of the model and provides insight into its suitability for efficient clinical deployment. Lower GMAC values correspond to faster execution and reduced hardware demands, allowing for direct comparison of computational efficiency across models while accounting for the fundamental arithmetic operations involved in neural network inference.
	\end{enumerate}
	\section{Results}
	This section describes the experimental results comparing the proposed HARU-Net against SOTA denoising methods. We first present quantitative performance metrics and then provide a visual analysis.
	
	\subsection{Quantitative Assessment}
	Table \ref{table1} lists the PSNR, SSIM, and GMSD values for each comparative method when applied to the testing data. Table \ref{table2} presents the computational cost metrics incurred during inference, expressed as the number of GMACs required to process one patch and the time (in seconds) required to process a complete 3D CBCT scan of size $512\times512\times512$. Bold values in each table indicate the best results. A combined view of the results in Tables \ref{table1} and \ref{table2} is presented in Fig. \ref{fig03} to demonstrate the balance between image fidelity and computational efficiency for each method.
	
	The results demonstrate that the proposed HARU-Net delivers the strongest overall performance, achieving the highest PSNR of 37.52 dB, the second-highest SSIM of 0.9557 (i.e., slightly below HAT, which achieves 0.9569), and the lowest GMSD (0.1084) among all comparative methods. These results indicate that HARU-Net provides the best balance between noise suppression and preservation of anatomical detail. Despite its superior performance, HARU-Net maintains a computational cost substantially lower than both Uformer and SwinIR. This trend is clearly illustrated in Fig. \ref{fig03}, which visualizes how PSNR and SSIM performance scale with respect to computational cost. ResU-Net achieves the lowest PSNR and highest GMSD among the compared models, although it demonstrates higher structural similarity to the ground truth than Uformer. Notably, ResU-Net is significantly faster than the transformer-based architectures.
	
	\subsection{Visual Assessment}
\begin{figure*}
		\centering
		\begin{subfigure}[b]{0.165\textwidth}
			\includegraphics[width=\textwidth]{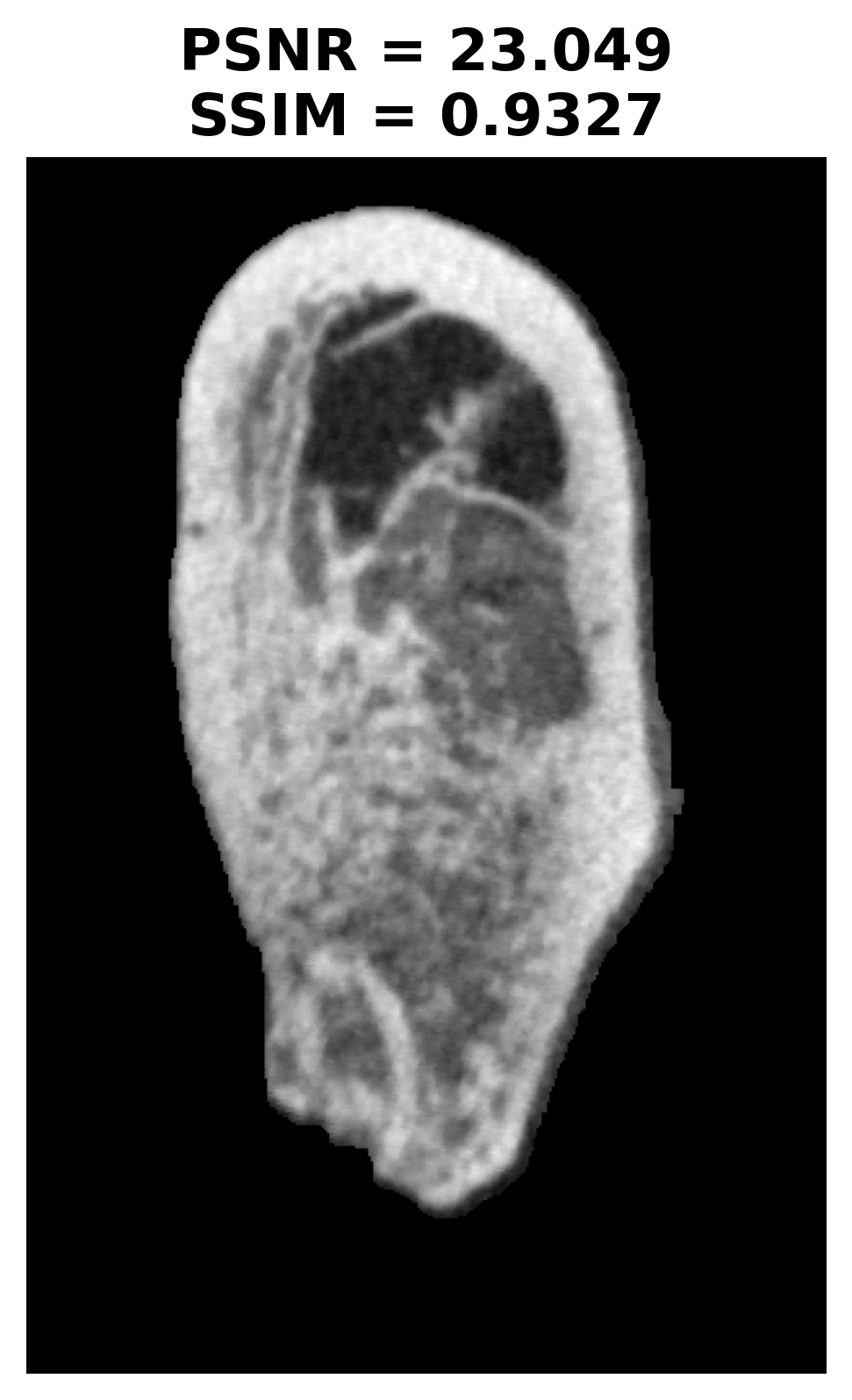}
		\end{subfigure}
		\begin{subfigure}[b]{0.165\textwidth}
			\includegraphics[width=\textwidth]{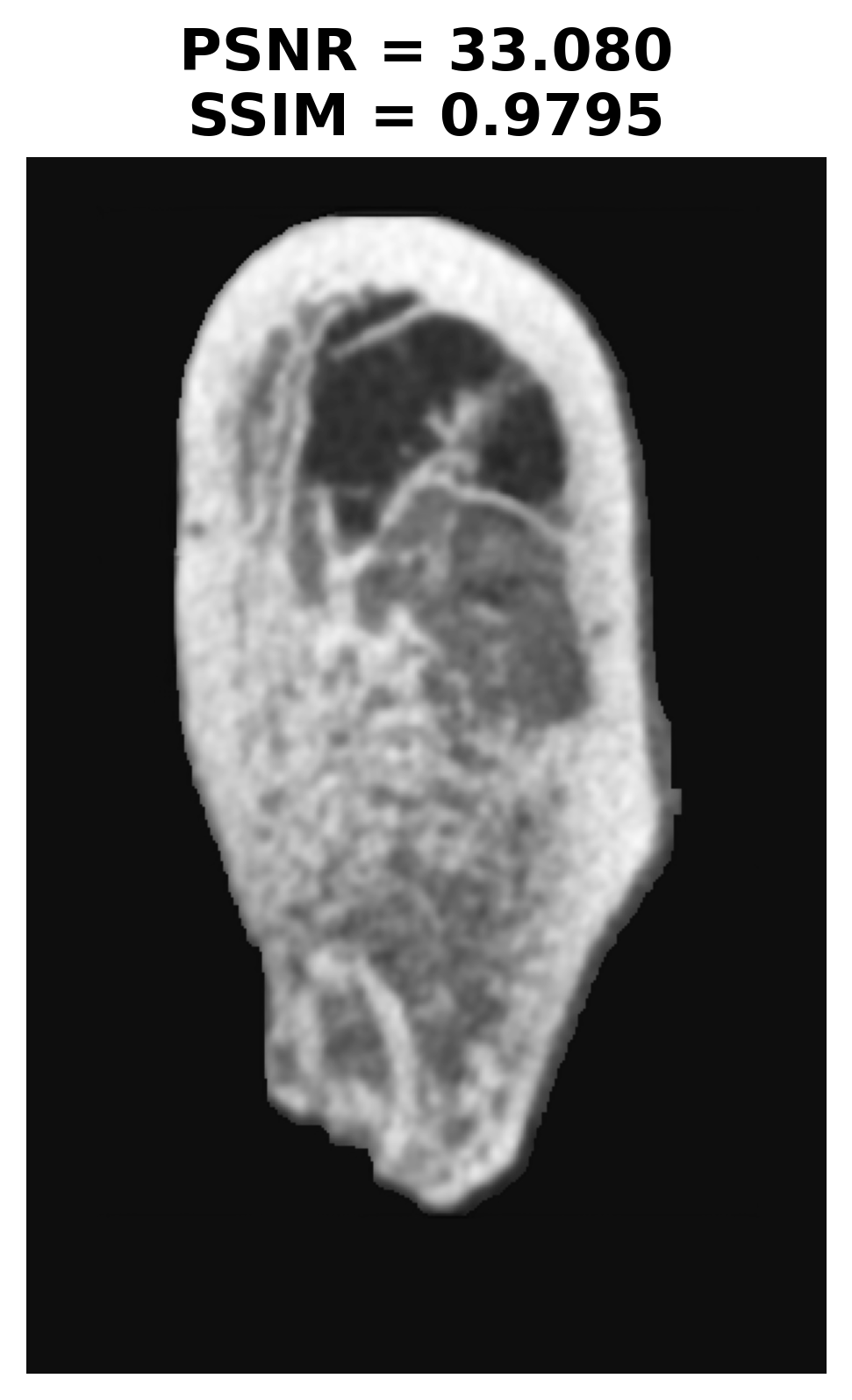}
		\end{subfigure}
		\hspace{-2.5mm}
		\begin{subfigure}[b]{0.165\textwidth}
			\includegraphics[width=\textwidth]{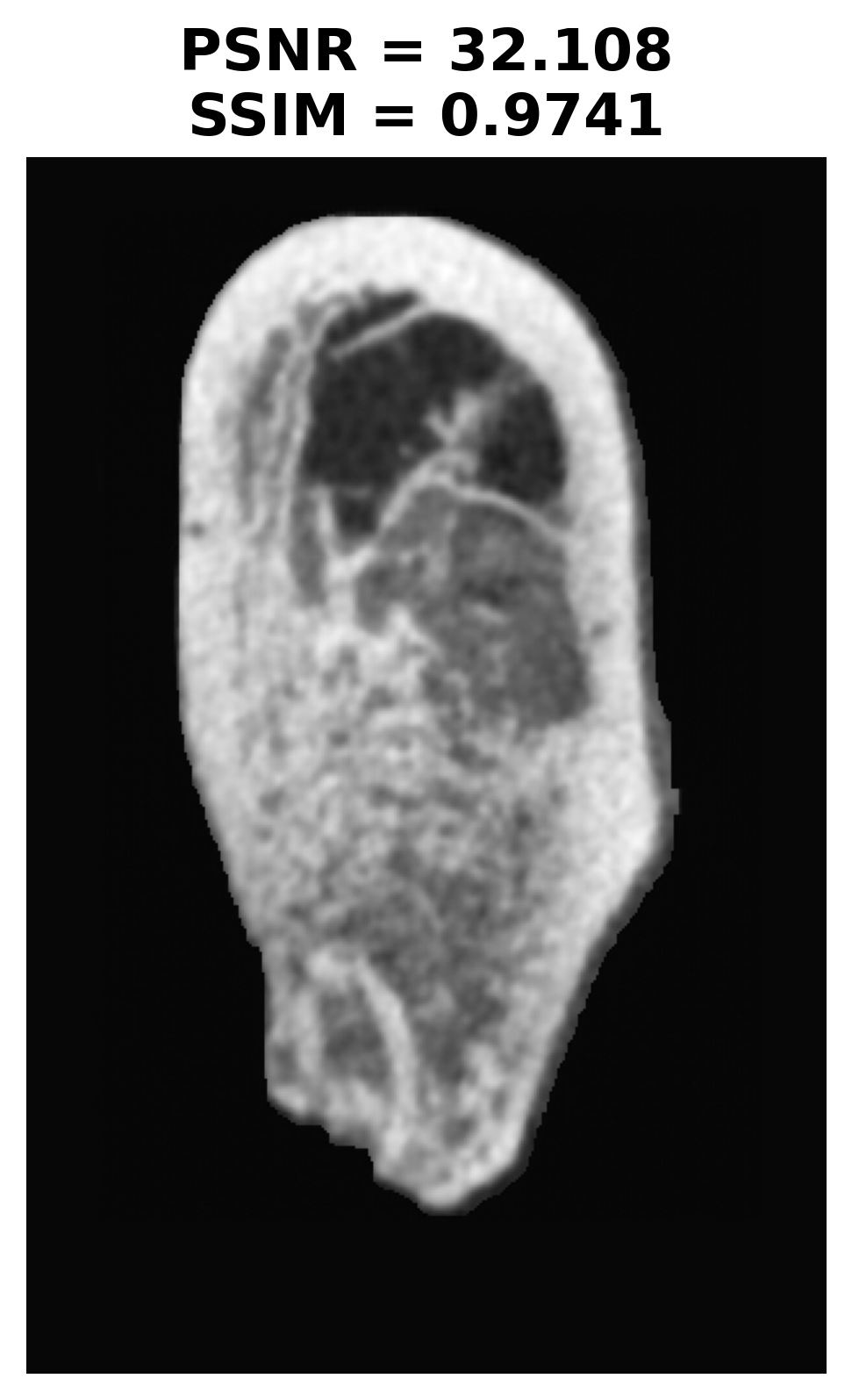}
		\end{subfigure}
		\hspace{-2.5mm}
		\begin{subfigure}[b]{0.165\textwidth}
			\includegraphics[width=\textwidth]{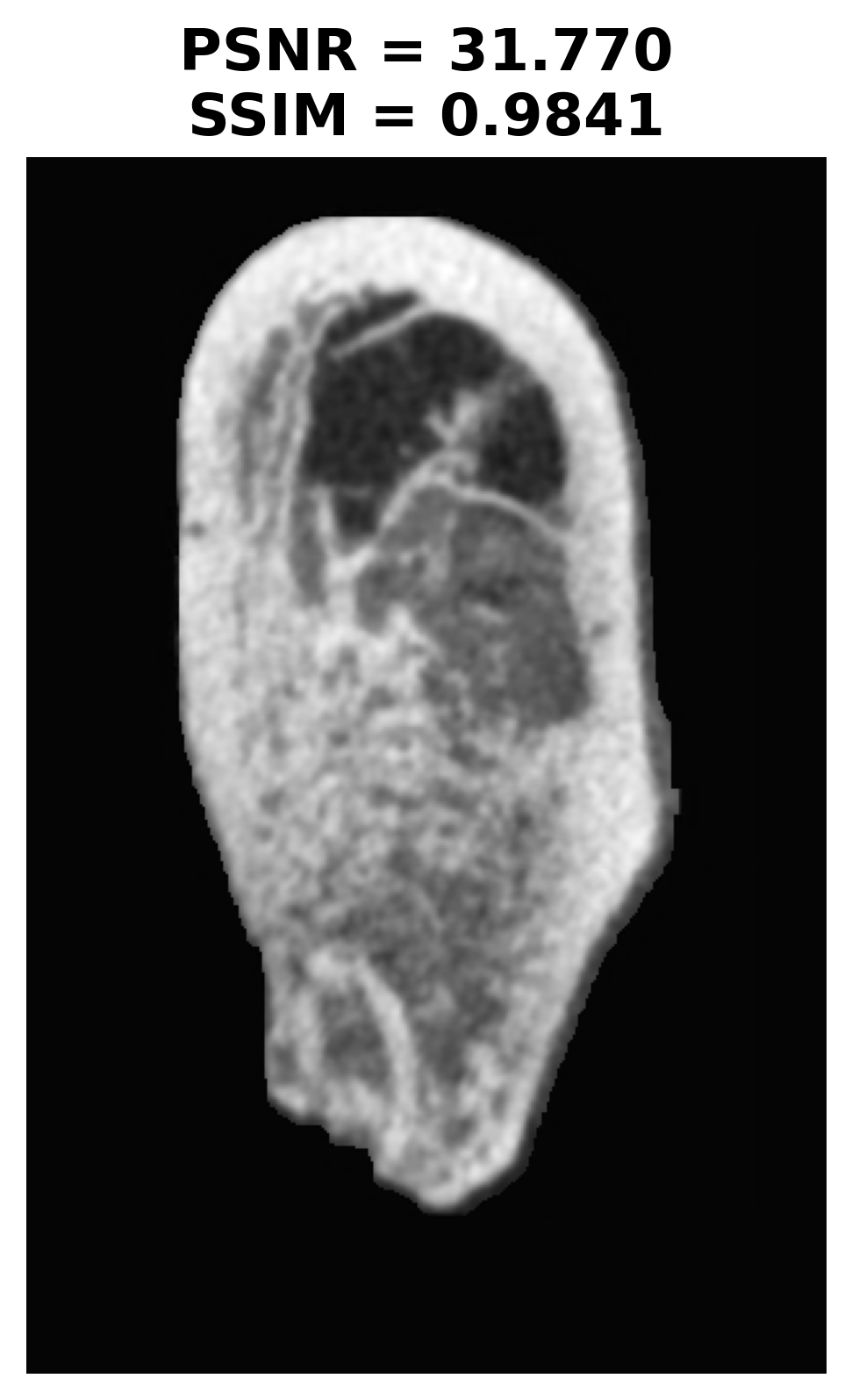}
		\end{subfigure}
		\hspace{-2.5mm}
		\begin{subfigure}[b]{0.165\textwidth}
			\includegraphics[width=\textwidth]{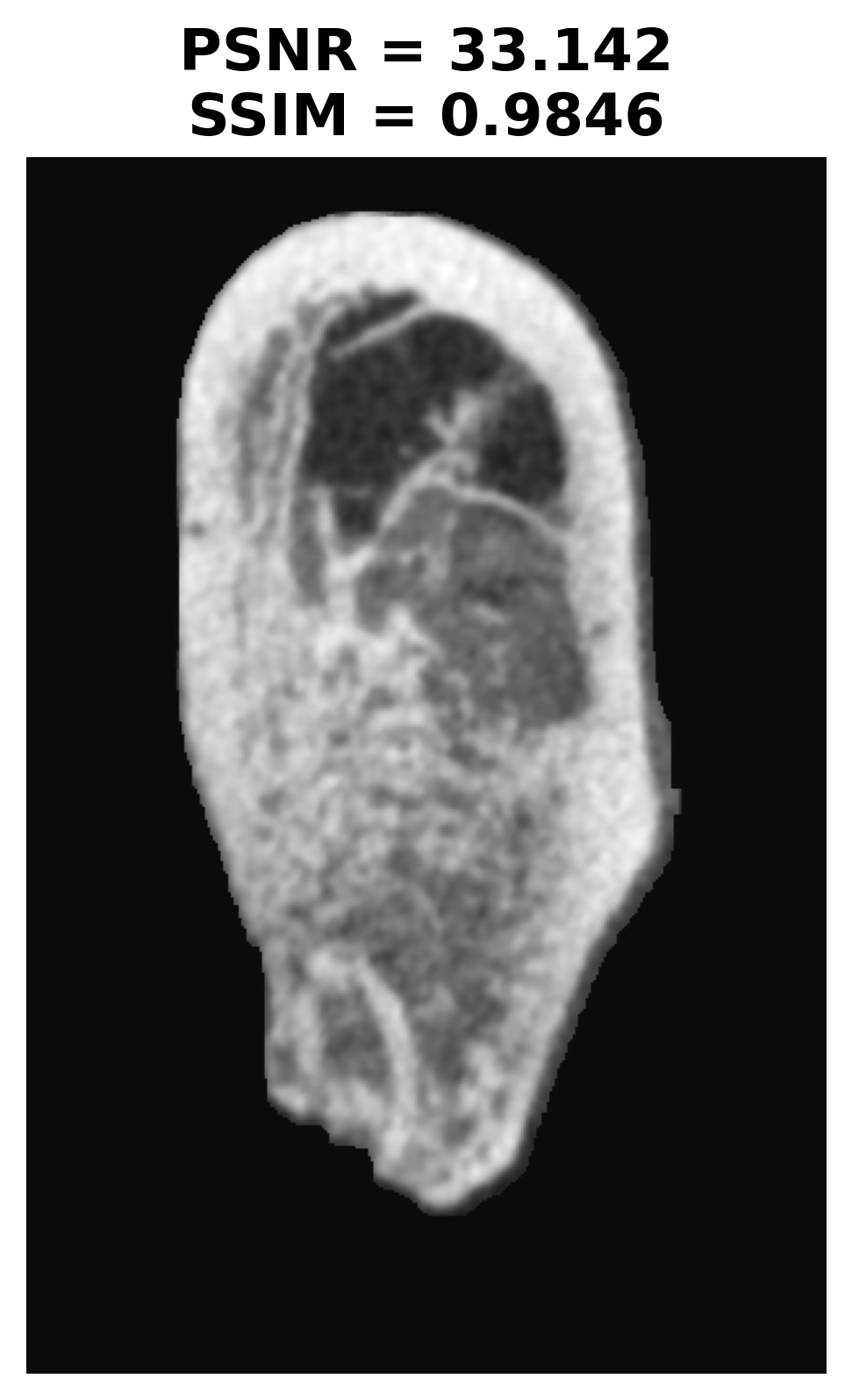}
		\end{subfigure}
		\hspace{-2.5mm}
		\begin{subfigure}[b]{0.165\textwidth}
			\includegraphics[width=\textwidth]{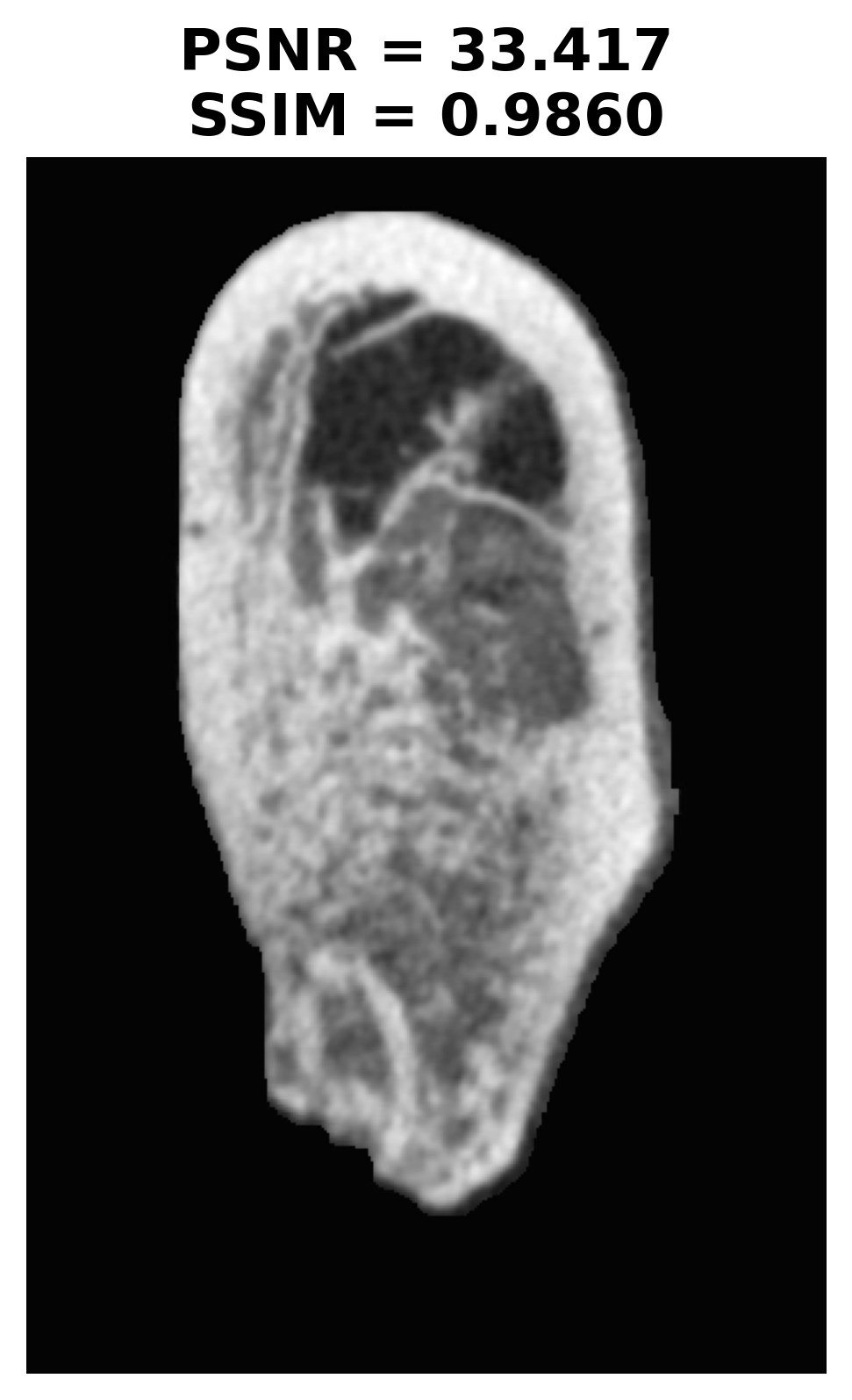}
		\end{subfigure}
		
		\begin{subfigure}[b]{0.1635\textwidth}
			\includegraphics[width=\textwidth]{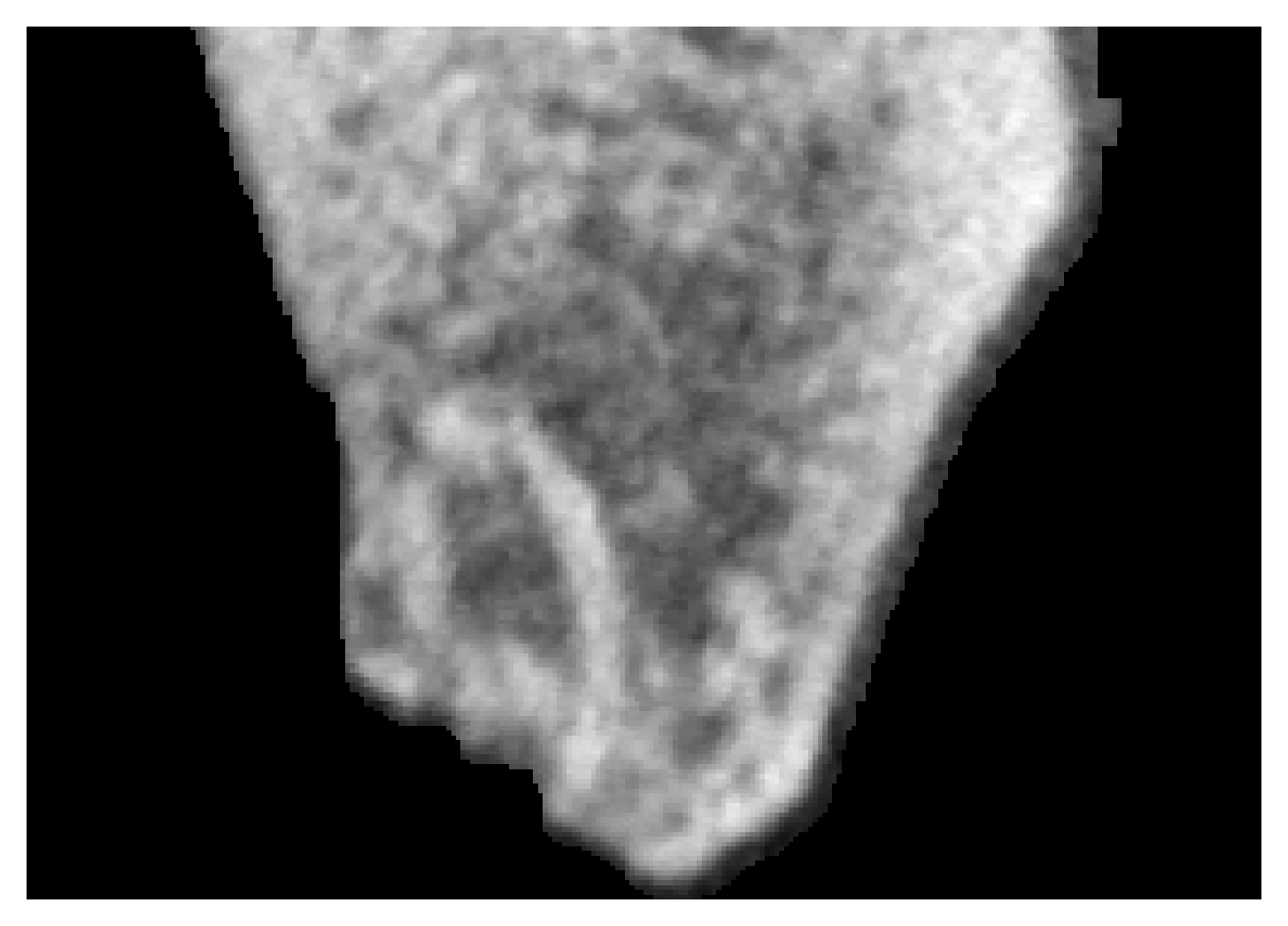}
			\caption{Noisy}
			\label{fig4a}
		\end{subfigure}
		\hspace{-2mm}
		\begin{subfigure}[b]{0.1635\textwidth}
			\includegraphics[width=\textwidth]{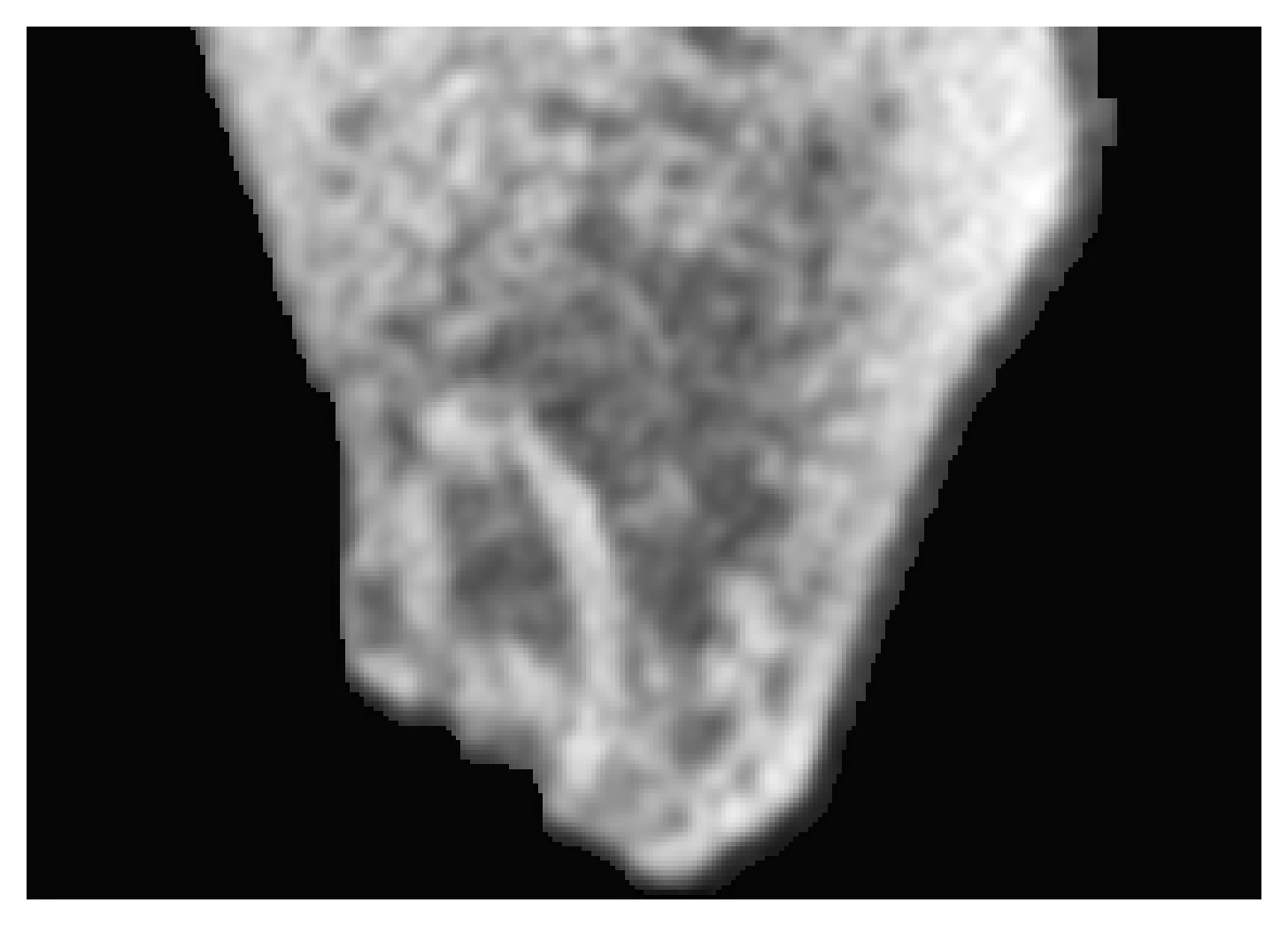}
			\caption{ResU-Net}
			\label{fig4b}
		\end{subfigure}
		\hspace{-2mm}
		\begin{subfigure}[b]{0.1635\textwidth}
			\includegraphics[width=\textwidth]{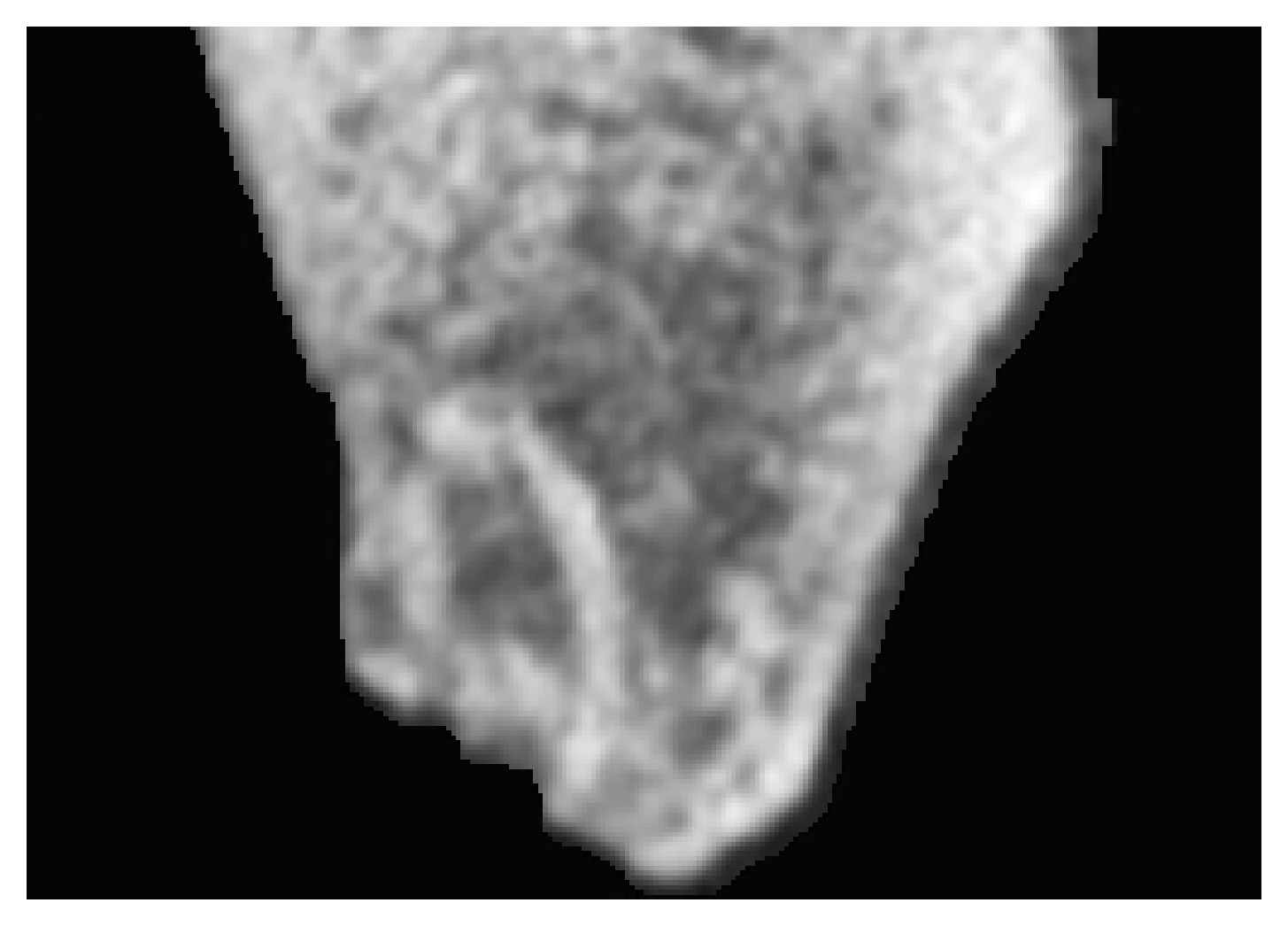}
			\caption{Uformer}
			\label{fig4c}
		\end{subfigure}
		\hspace{-2mm}
		\begin{subfigure}[b]{0.1635\textwidth}
			\includegraphics[width=\textwidth]{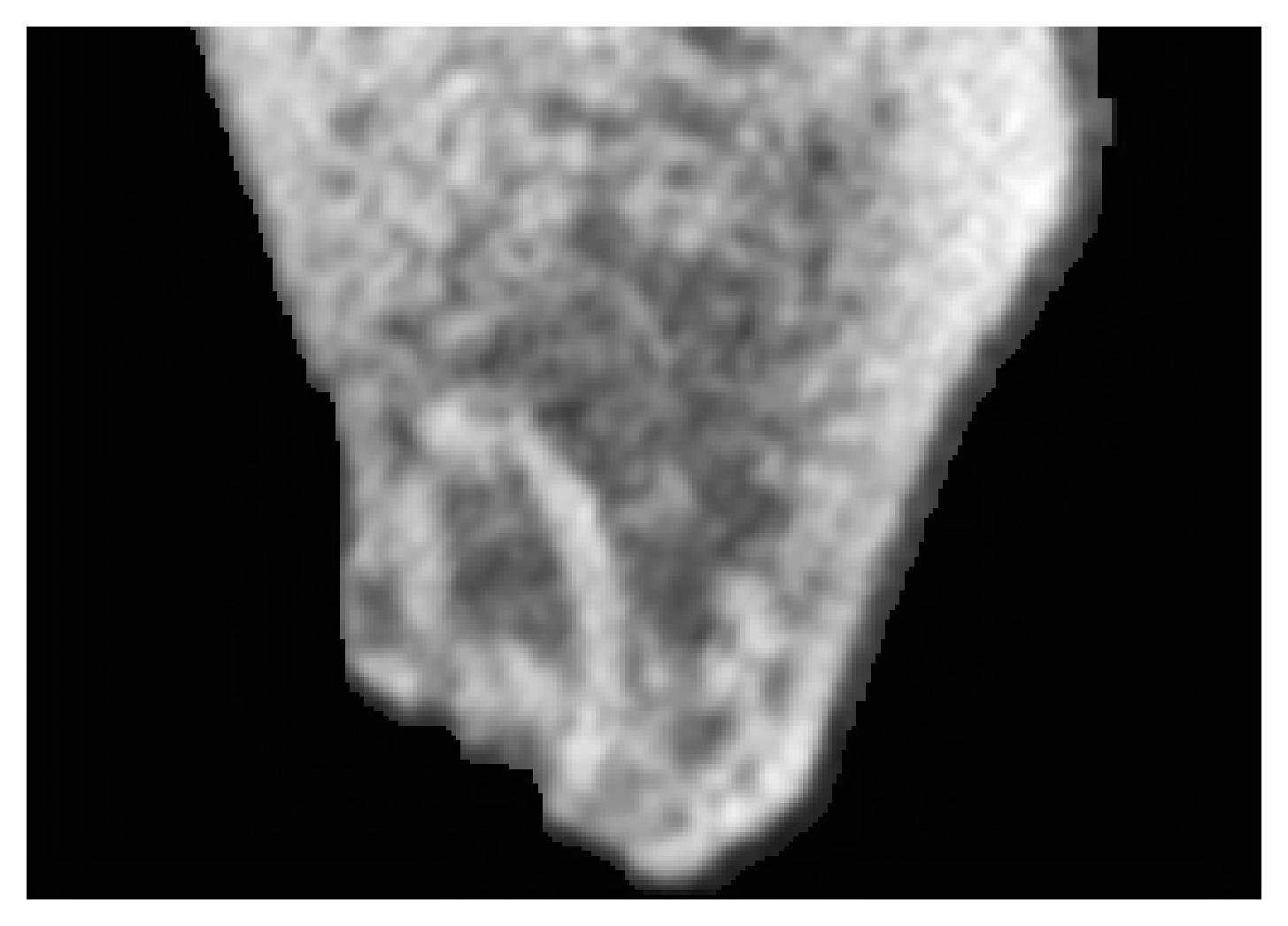}
			\caption{SwinIR}
			\label{fig4b}
		\end{subfigure}
		\hspace{-2mm}
		\begin{subfigure}[b]{0.1635\textwidth}
			\includegraphics[width=\textwidth]{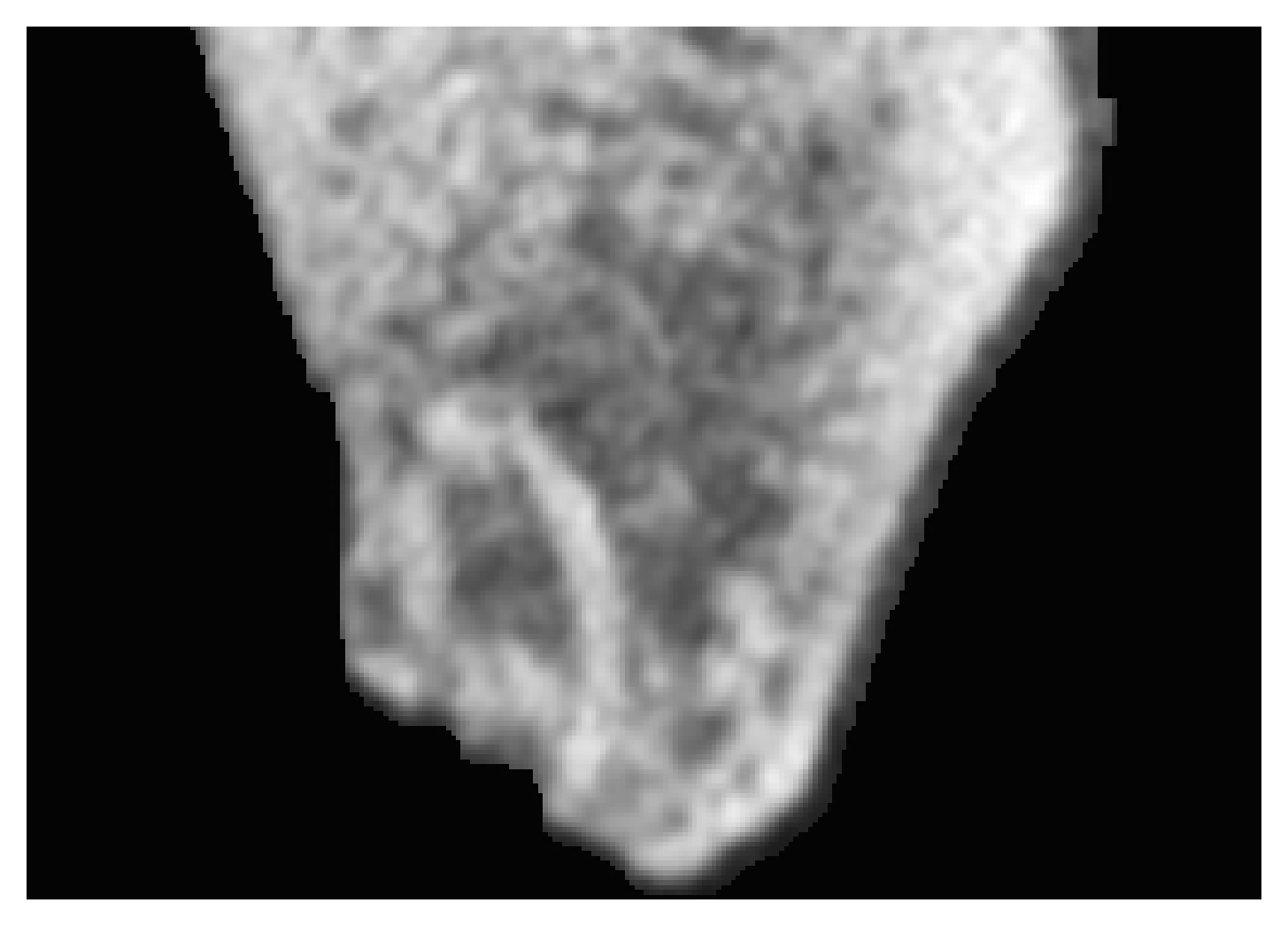}
			\caption{HAT}
			\label{fig4b}
		\end{subfigure}
		\hspace{-2mm}     
		\begin{subfigure}[b]{0.1635\textwidth}
			\includegraphics[width=\textwidth]{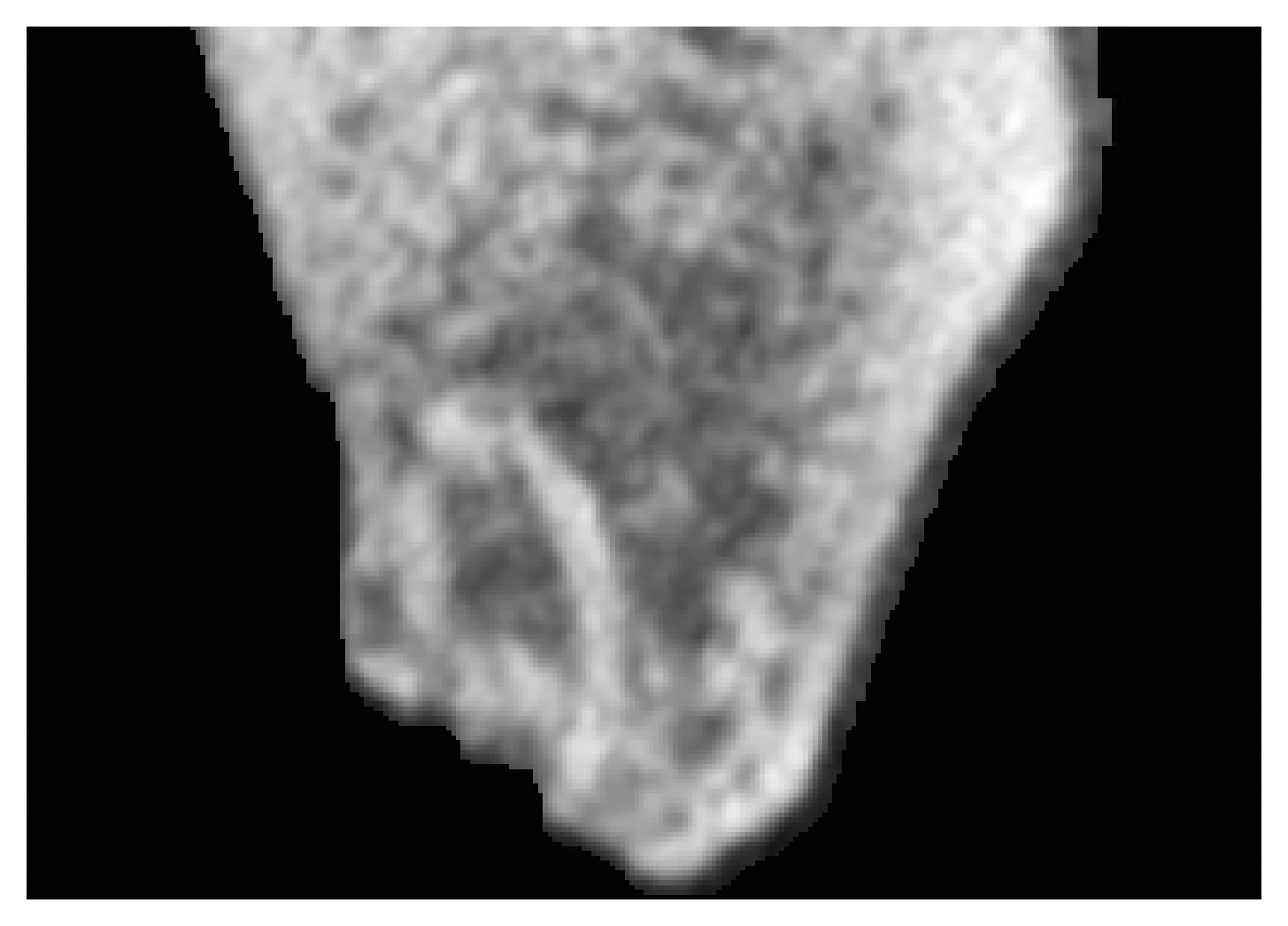}
			\caption{HARU-Net}
			\label{fig4b}
		\end{subfigure}
		\caption{Comparison of denoising performance on CBCT slice from sagital view from a 3D CBCT scan.}
		\label{fig04}
	\end{figure*}

\begin{figure*}
		\centering
		\begin{subfigure}[b]{0.17\textwidth}
			\includegraphics[width=\textwidth]{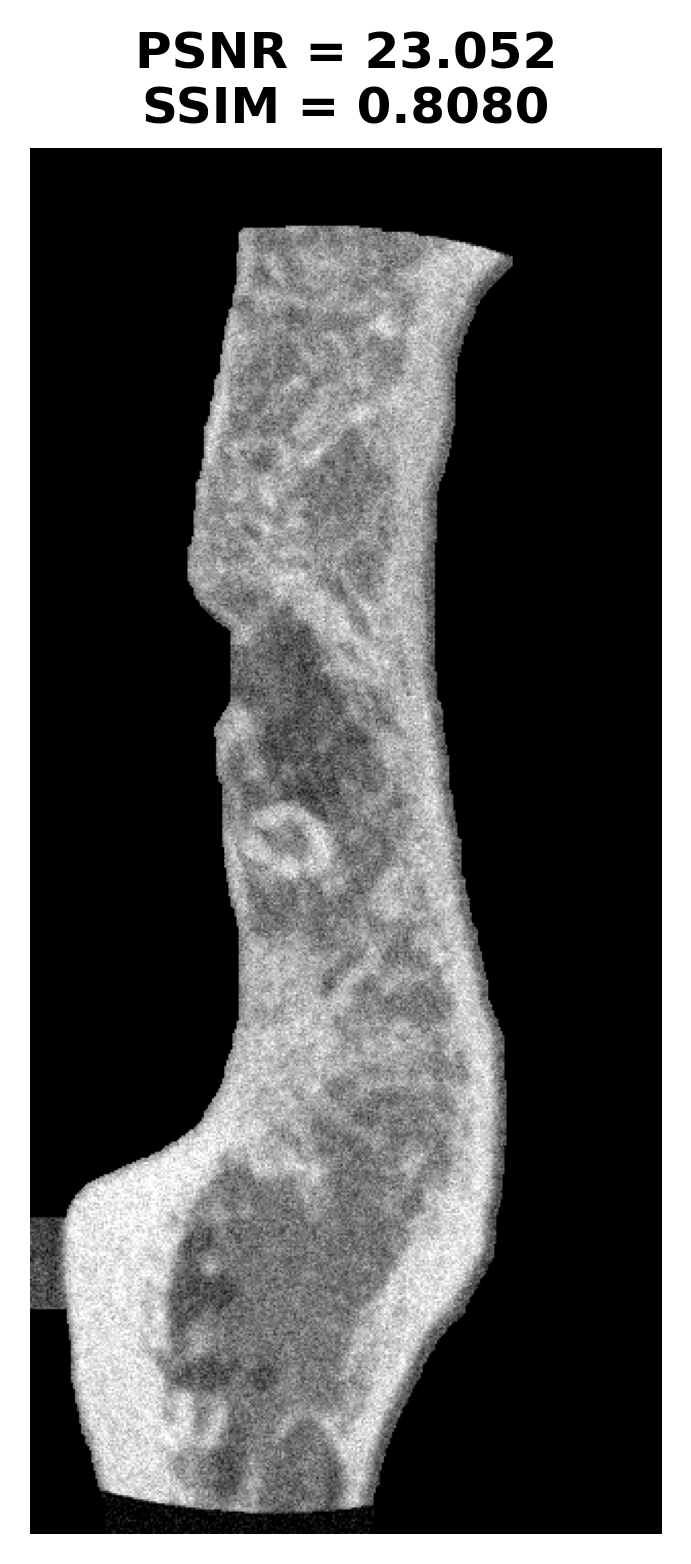}
		\end{subfigure}
		\hspace{-3mm}
		\begin{subfigure}[b]{0.17\textwidth}
			\includegraphics[width=\textwidth]{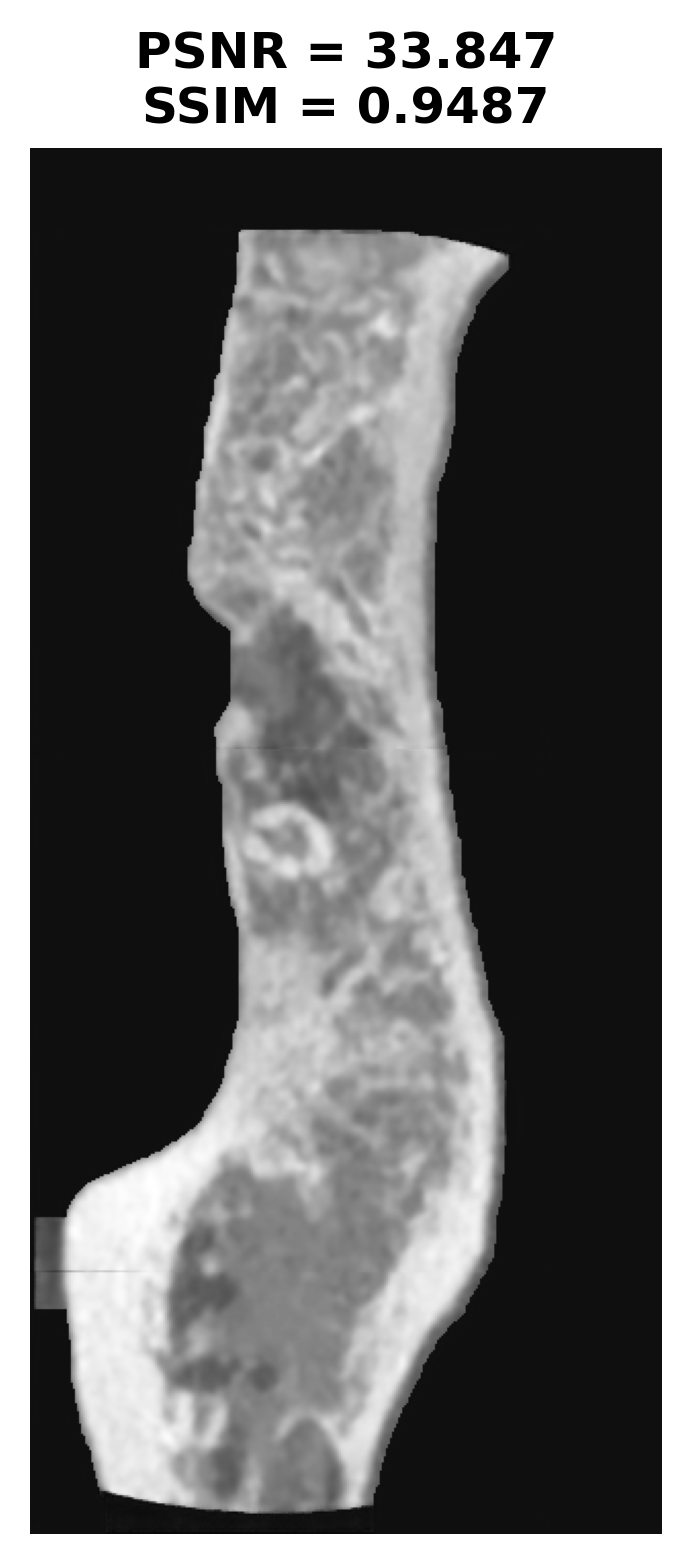}
		\end{subfigure}
		\hspace{-3mm}
		\begin{subfigure}[b]{0.17\textwidth}
			\includegraphics[width=\textwidth]{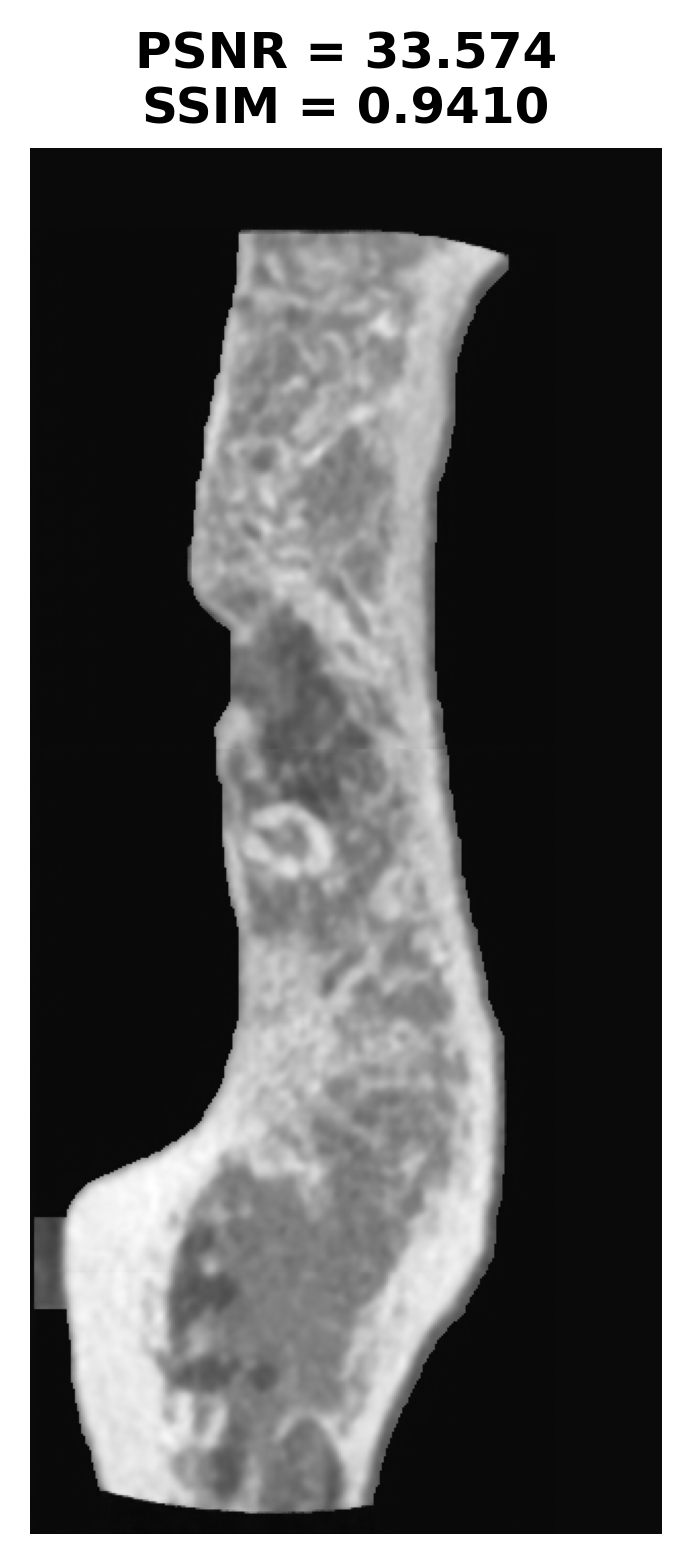}
		\end{subfigure}
		\hspace{-3mm}
		\begin{subfigure}[b]{0.17\textwidth}
			\includegraphics[width=\textwidth]{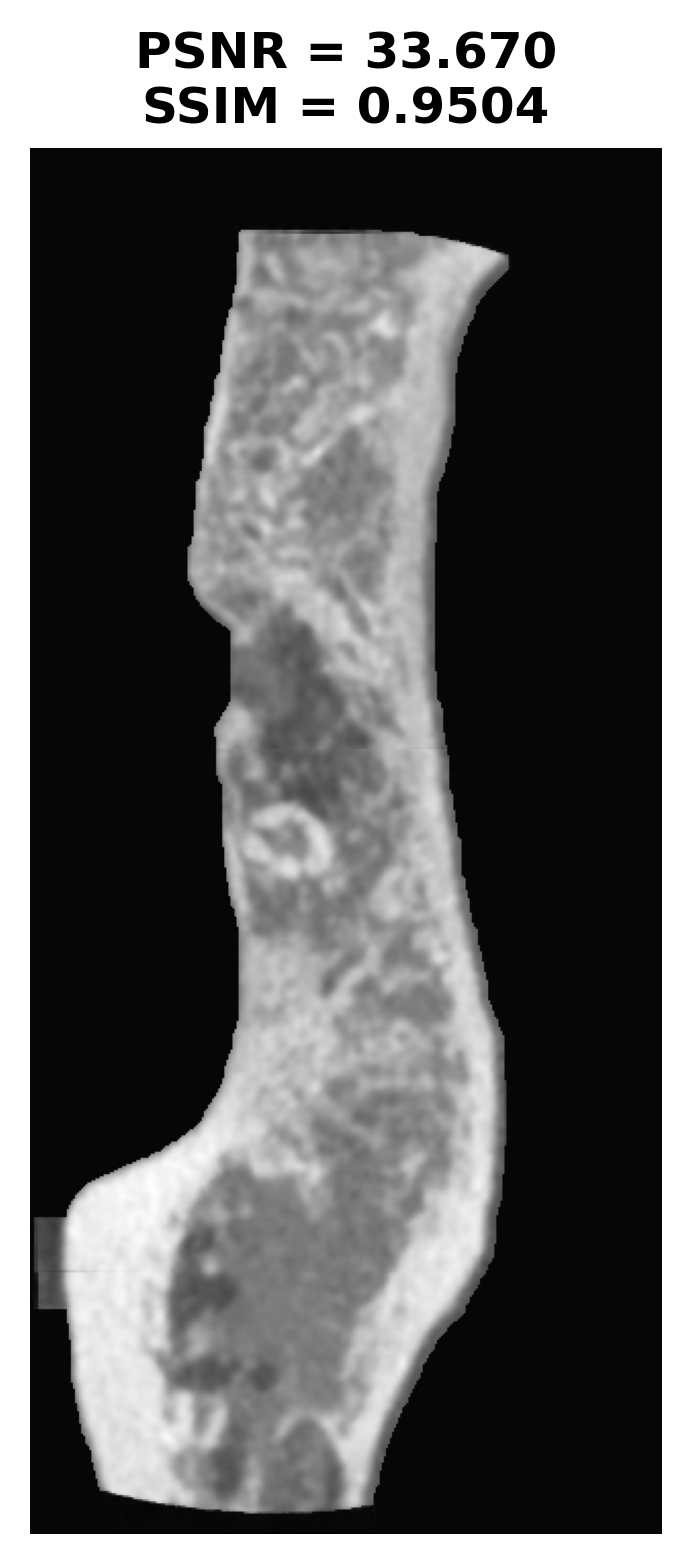}
		\end{subfigure}
		\hspace{-3mm}
		\begin{subfigure}[b]{0.17\textwidth}
			\includegraphics[width=\textwidth]{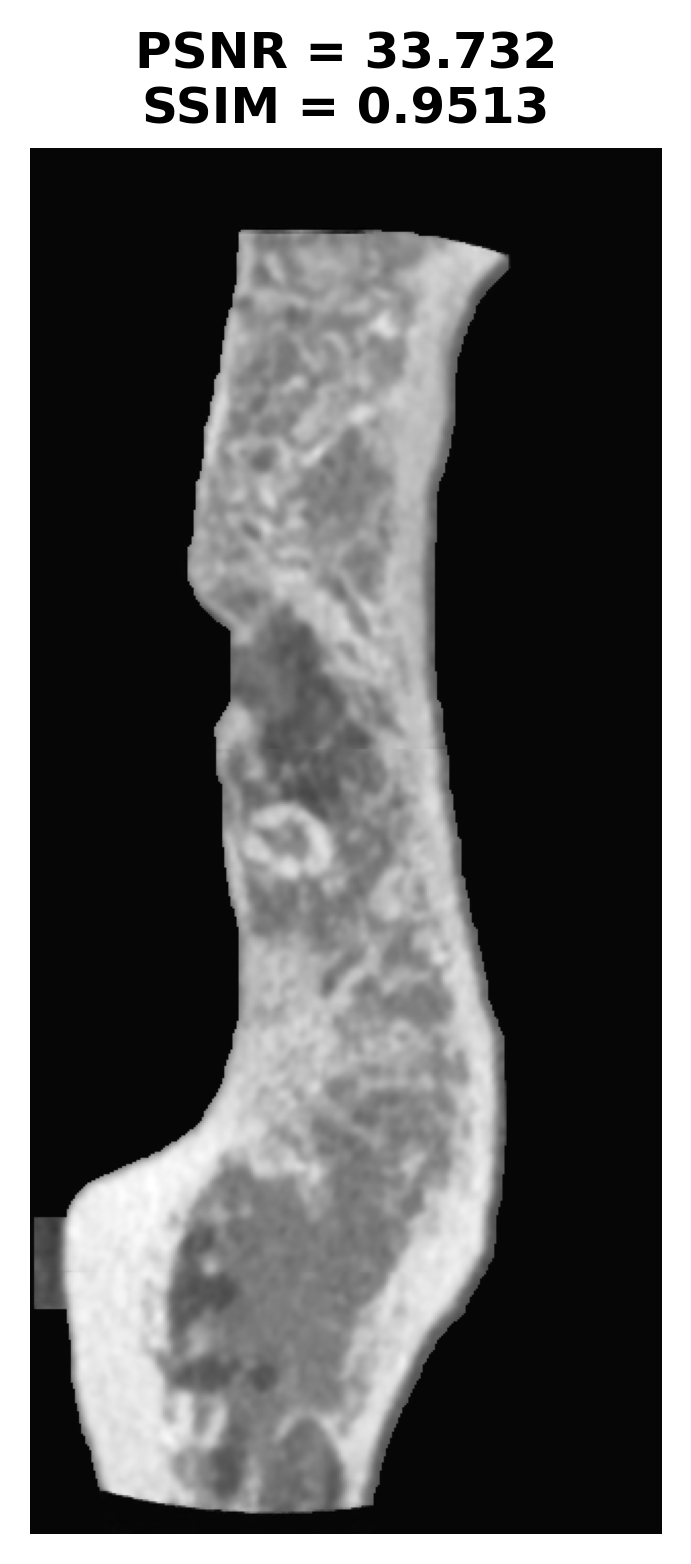}
		\end{subfigure}
		\hspace{-3mm}
		\begin{subfigure}[b]{0.17\textwidth}
			\includegraphics[width=\textwidth]{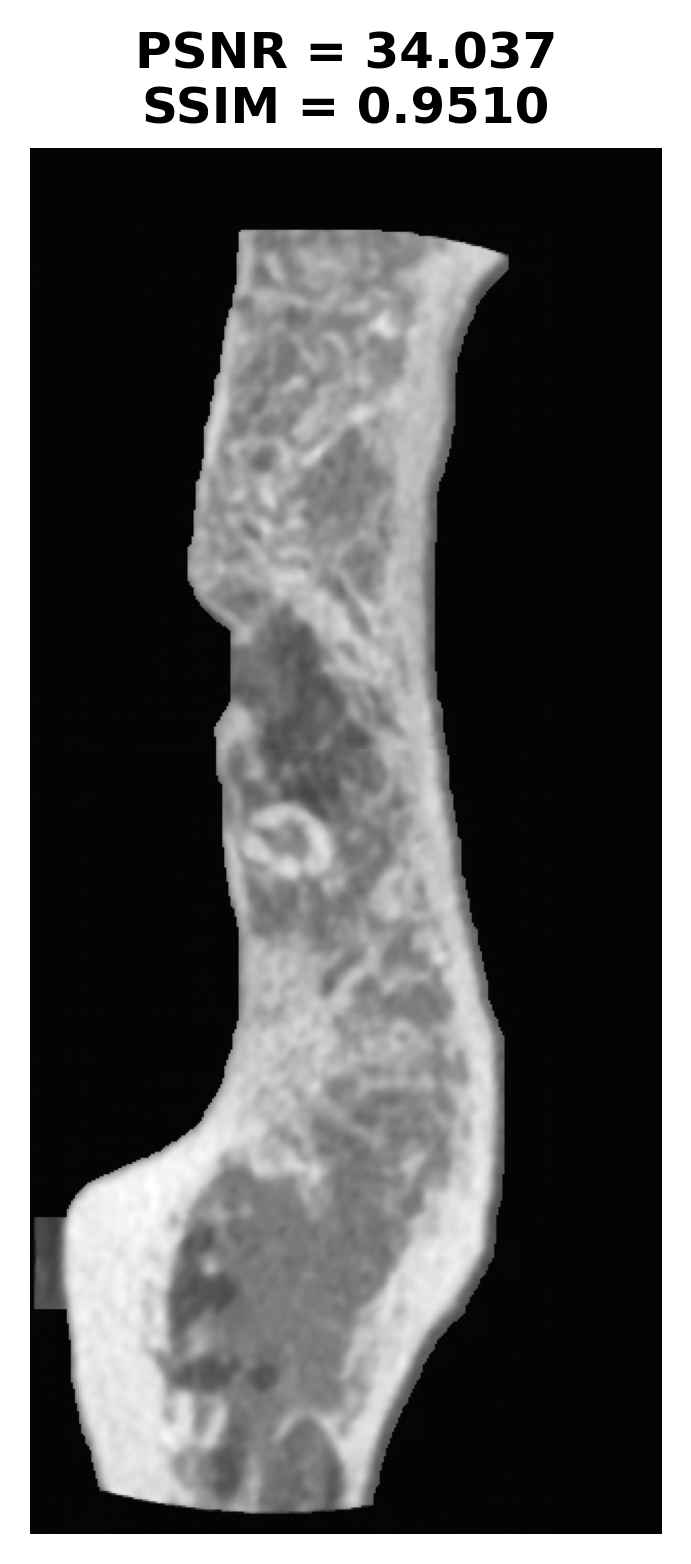}
		\end{subfigure}
		
		\vspace{-1mm}
		\begin{subfigure}[b]{0.163\textwidth}
			\includegraphics[width=\textwidth]{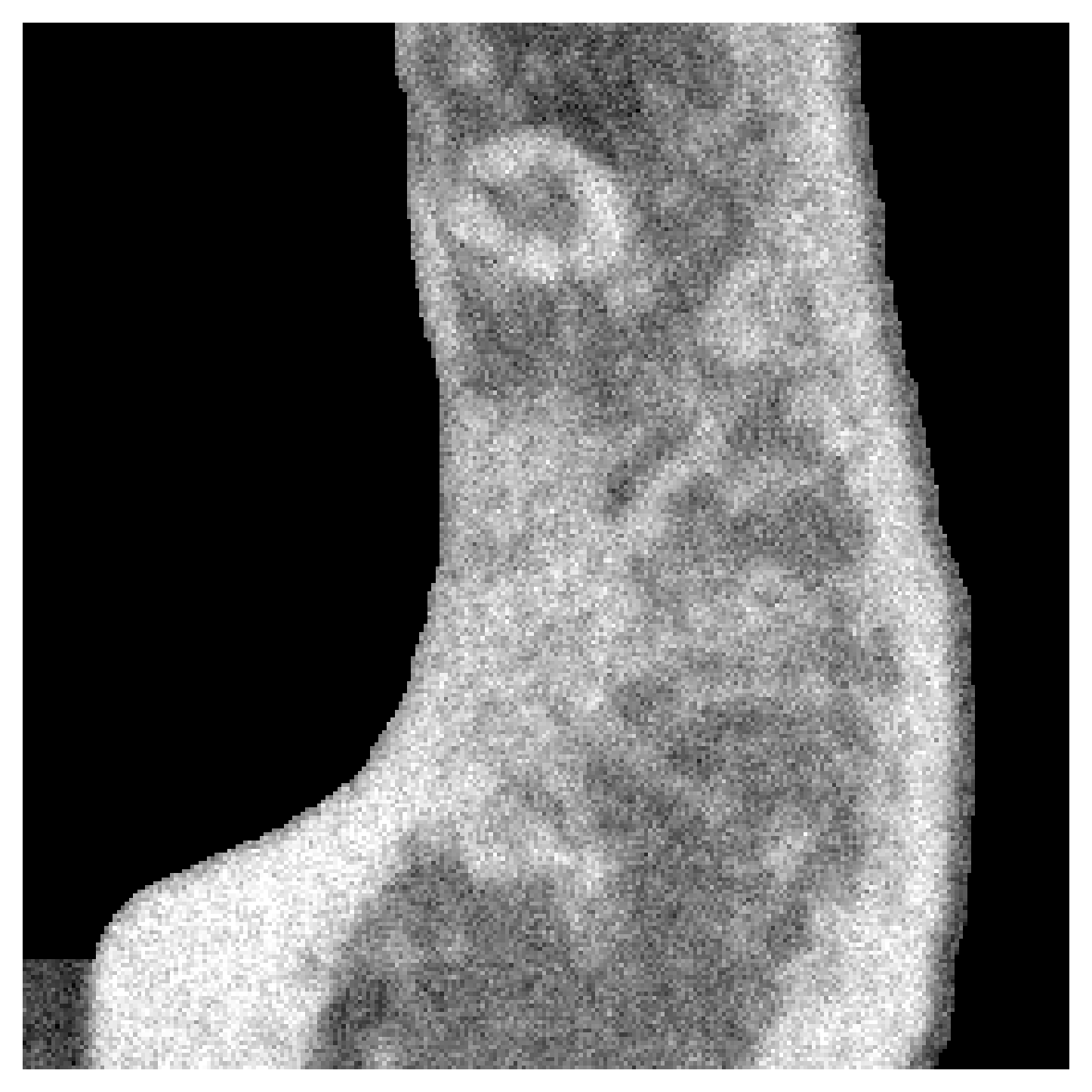}
			\caption{Noisy}
			\label{figVR1a}
		\end{subfigure}
		\hspace{-2mm}
		\begin{subfigure}[b]{0.163\textwidth}
			\includegraphics[width=\textwidth]{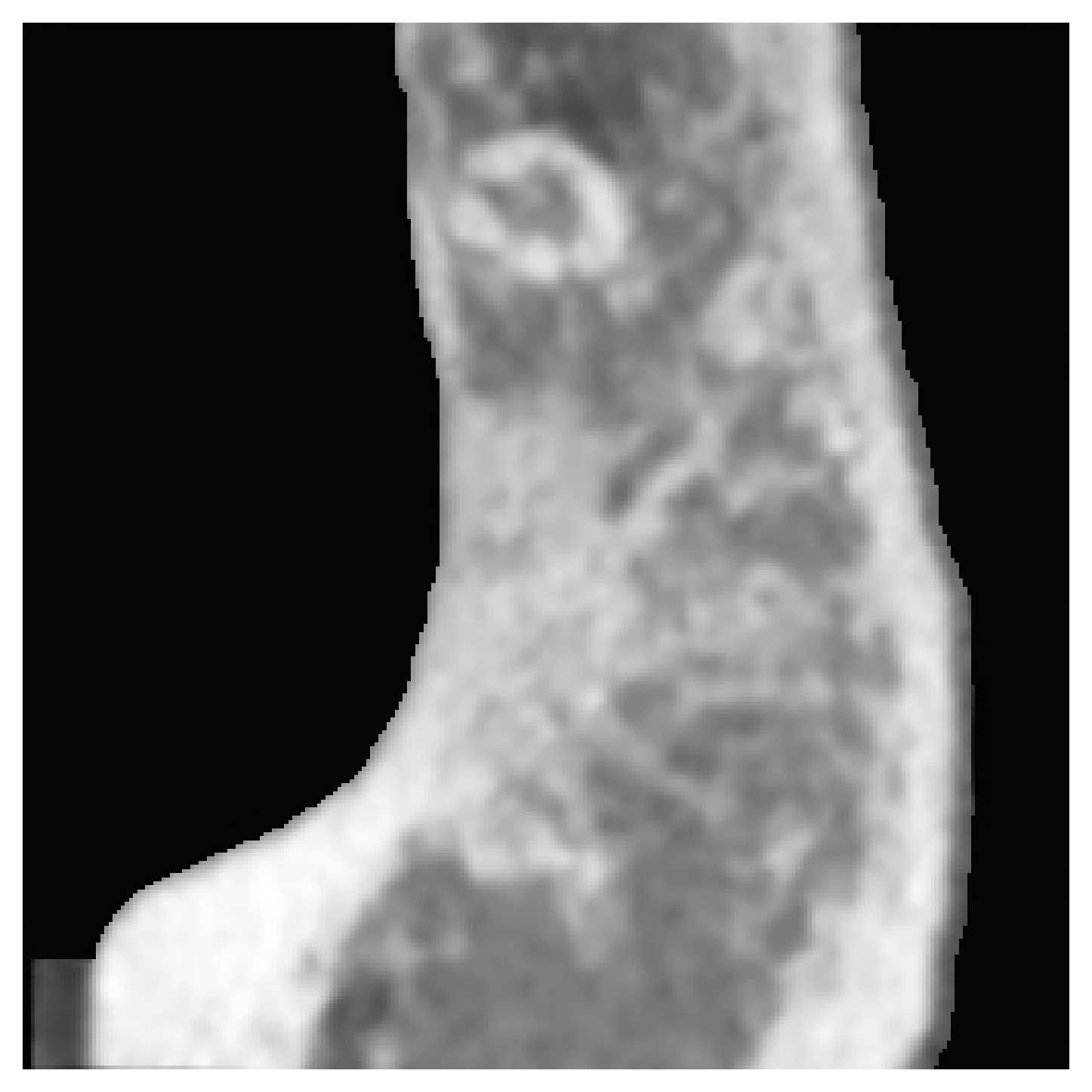}
			\caption{ResU-Net}
			\label{fig5b}
		\end{subfigure}
		\hspace{-2mm}
		\begin{subfigure}[b]{0.163\textwidth}
			\includegraphics[width=\textwidth]{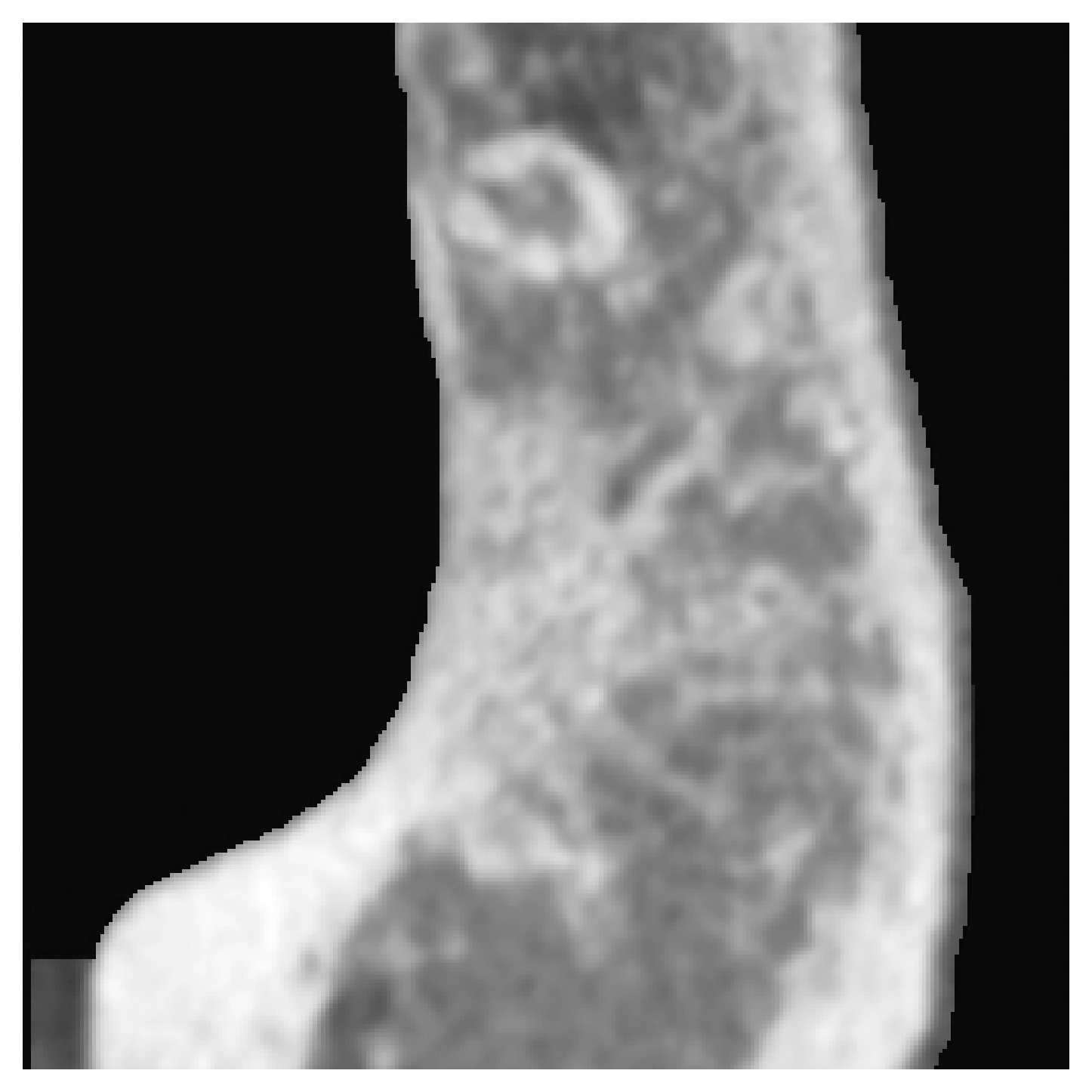}
			\caption{Uformer}
			\label{fig5c}
		\end{subfigure}
		\hspace{-2mm}
		\begin{subfigure}[b]{0.163\textwidth}
			\includegraphics[width=\textwidth]{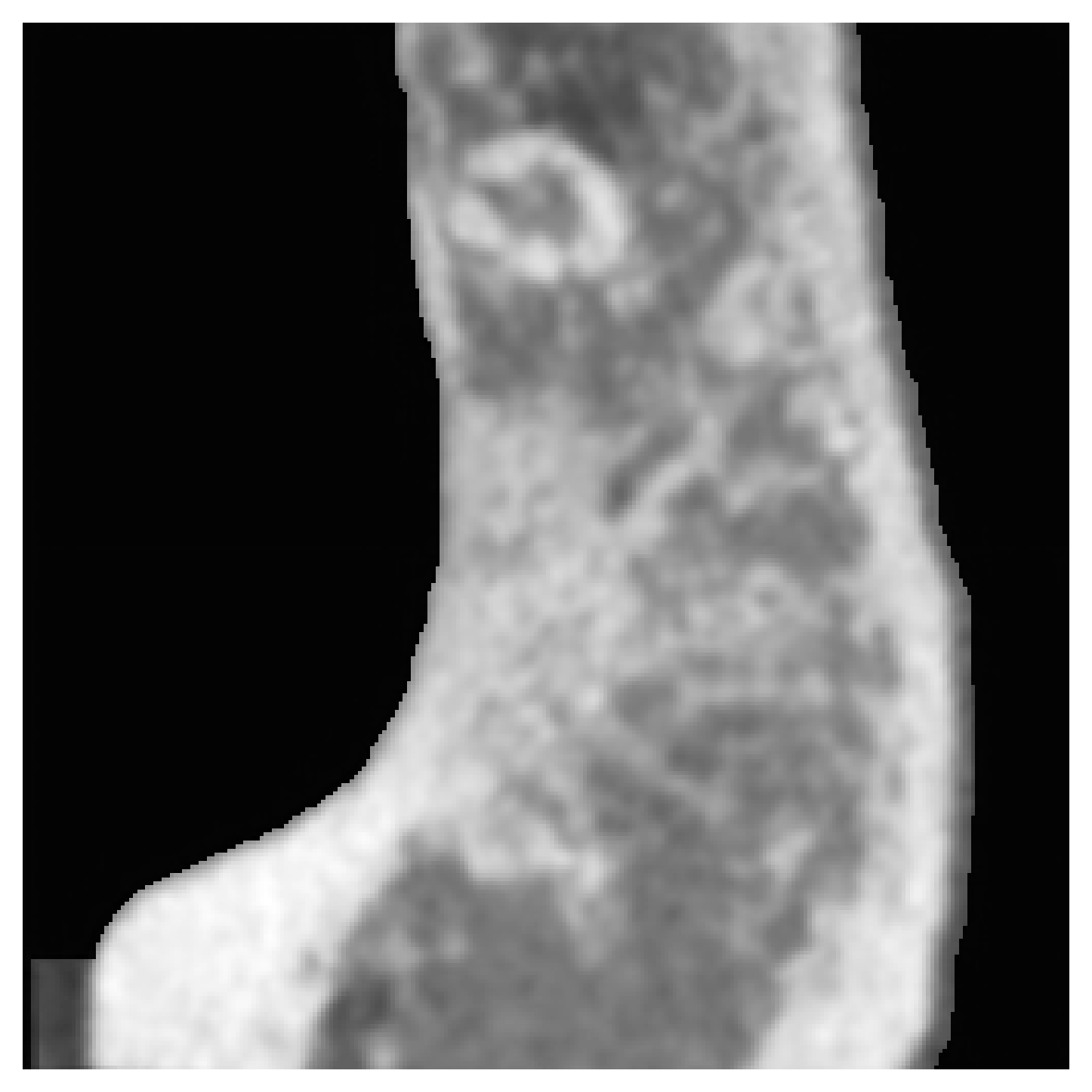}
			\caption{SwinIR}
			\label{fig5b}
		\end{subfigure}
		\hspace{-2mm}
		\begin{subfigure}[b]{0.163\textwidth}
			\includegraphics[width=\textwidth]{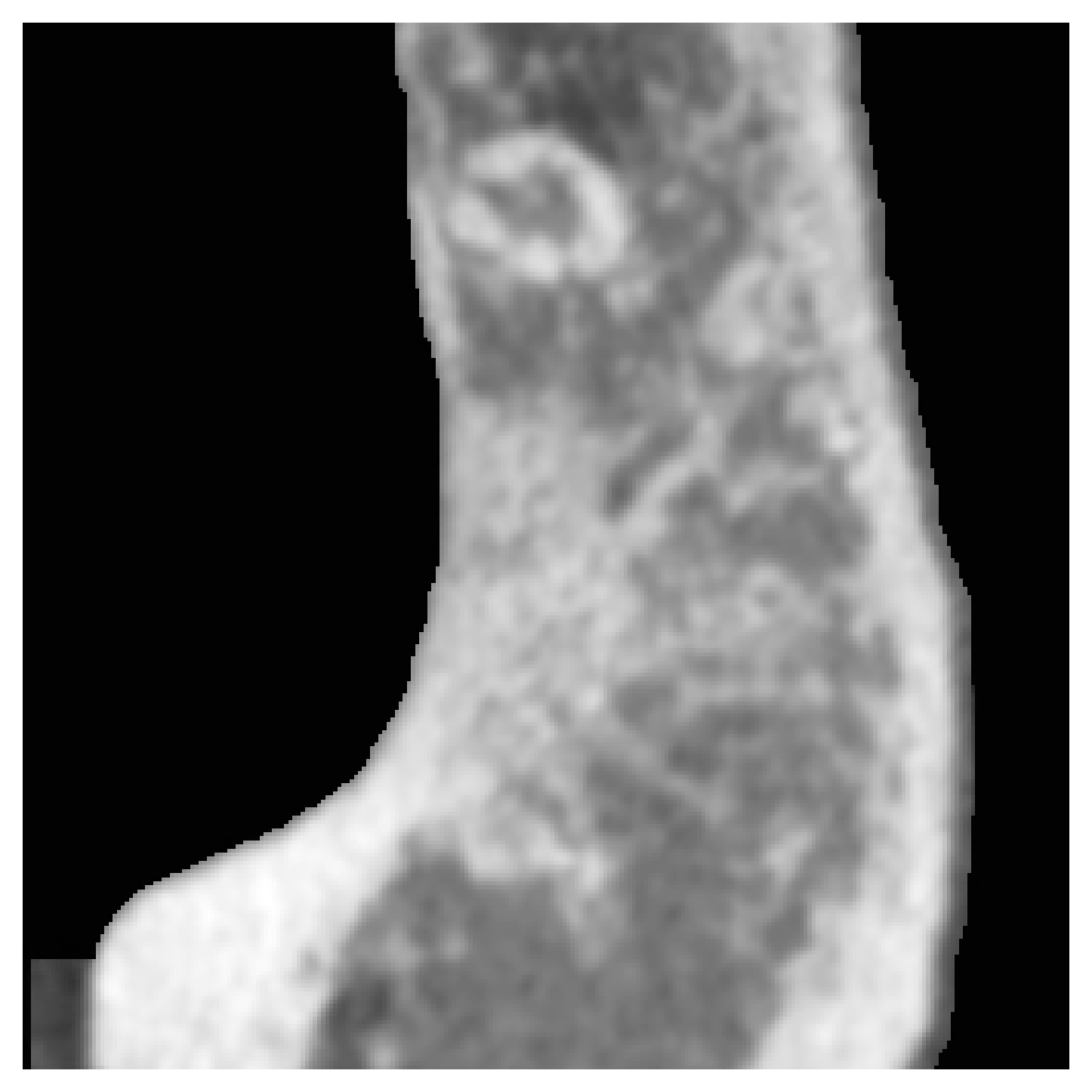}
			\caption{HAT}
			\label{fig5b}
		\end{subfigure}
		\hspace{-2mm}
		\begin{subfigure}[b]{0.163\textwidth}
			\includegraphics[width=\textwidth]{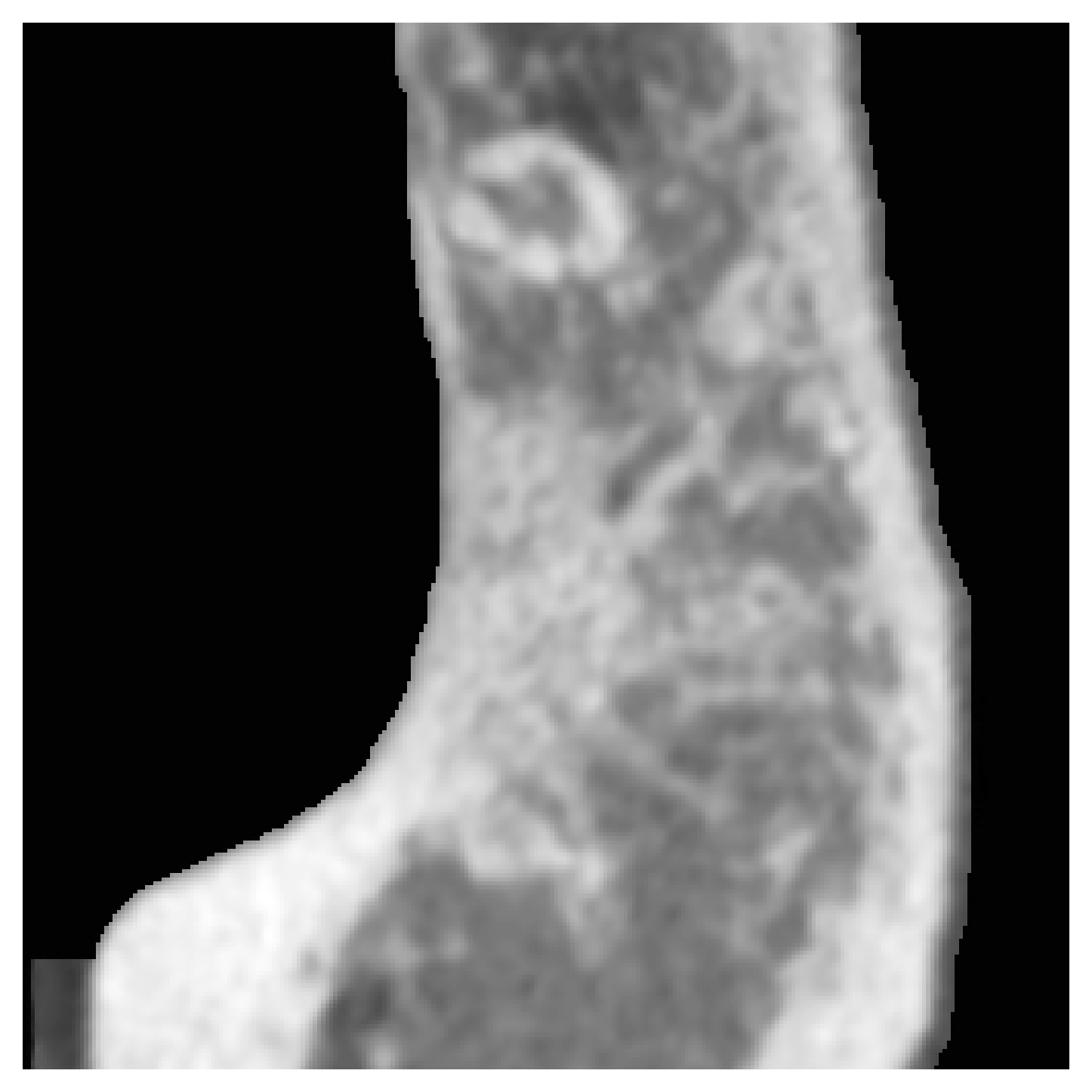}
			\caption{HARU-Net}
			\label{fig5b}
		\end{subfigure}
		\caption{Comparison of denoising performance on CBCT slice from axial view from a 3D CBCT scan.}
		\label{fig05}
	\end{figure*}

\begin{figure*}
		\centering
		\begin{subfigure}[b]{0.16\textwidth}
			\includegraphics[width=\textwidth]{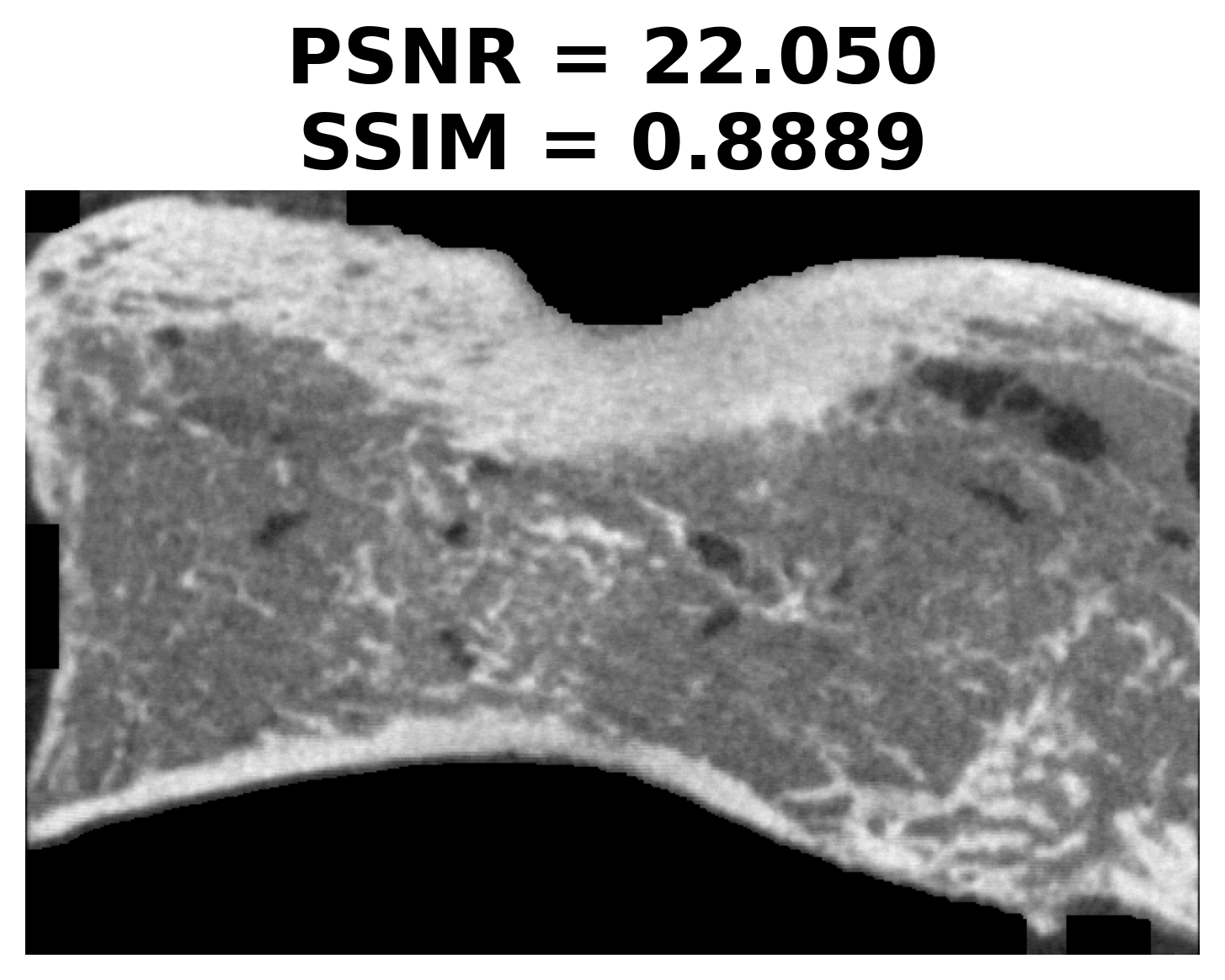}
		\end{subfigure}
		\begin{subfigure}[b]{0.16\textwidth}
			\includegraphics[width=\textwidth]{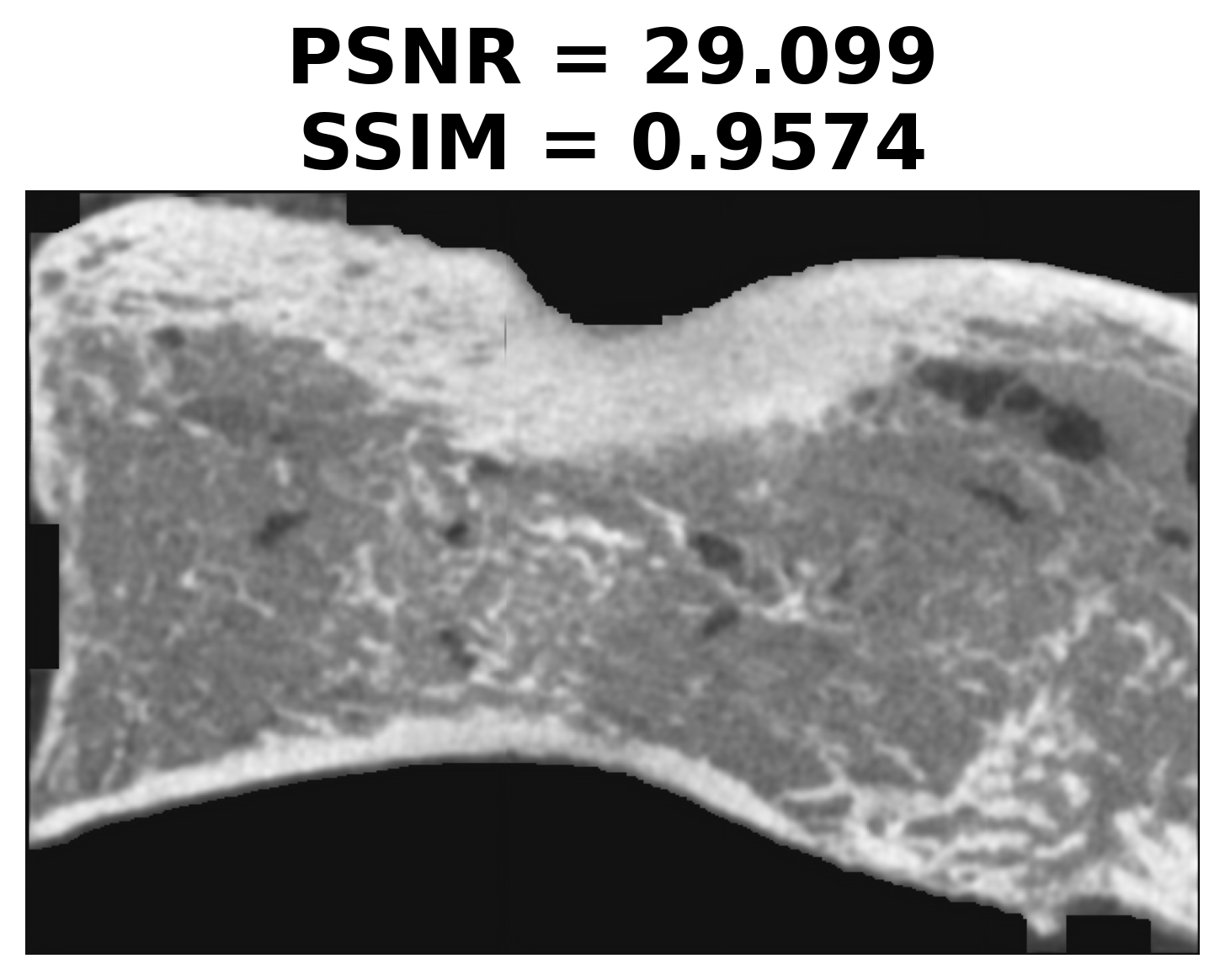}
		\end{subfigure}
		\begin{subfigure}[b]{0.16\textwidth}
			\includegraphics[width=\textwidth]{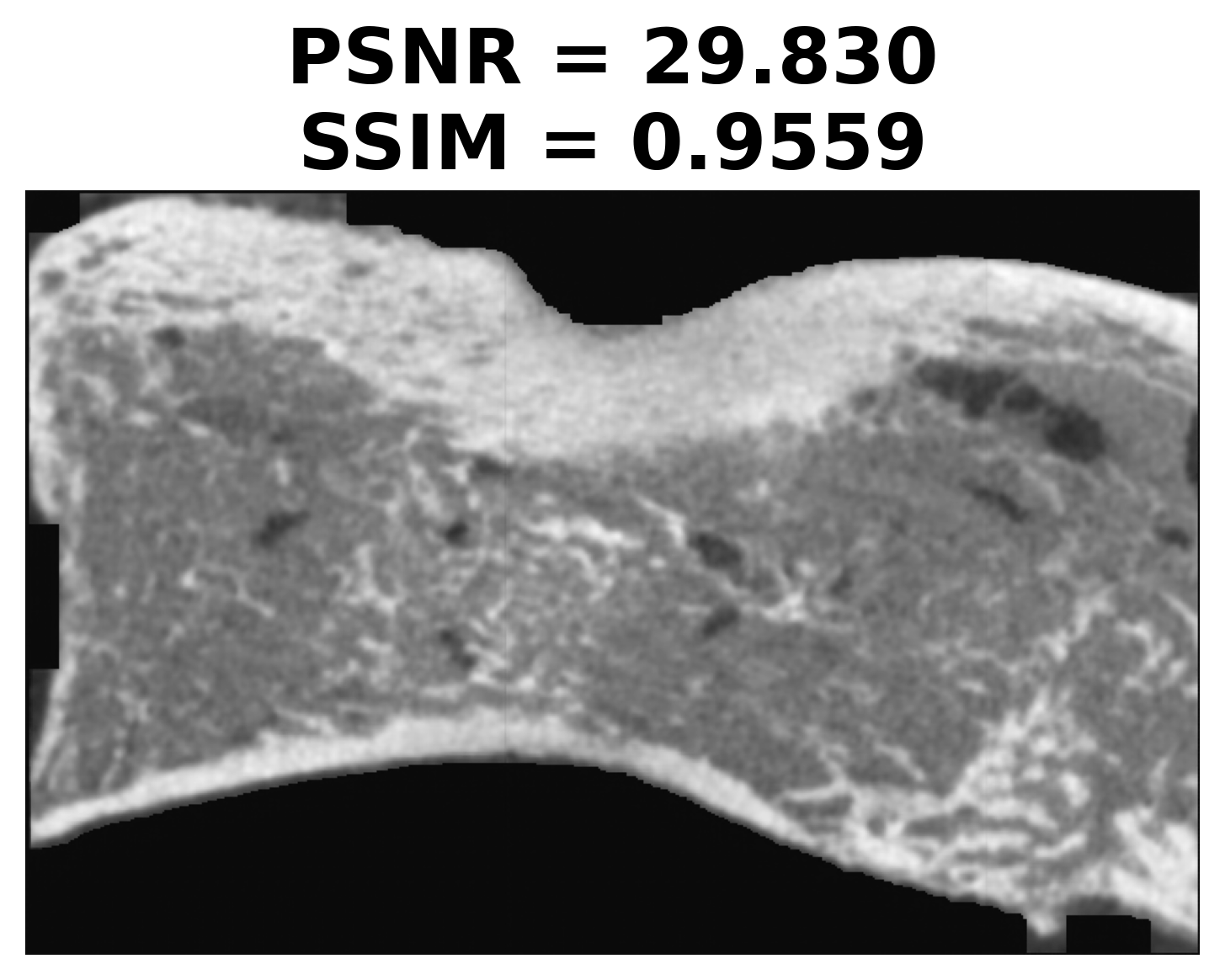}
		\end{subfigure}
		\begin{subfigure}[b]{0.16\textwidth}
			\includegraphics[width=\textwidth]{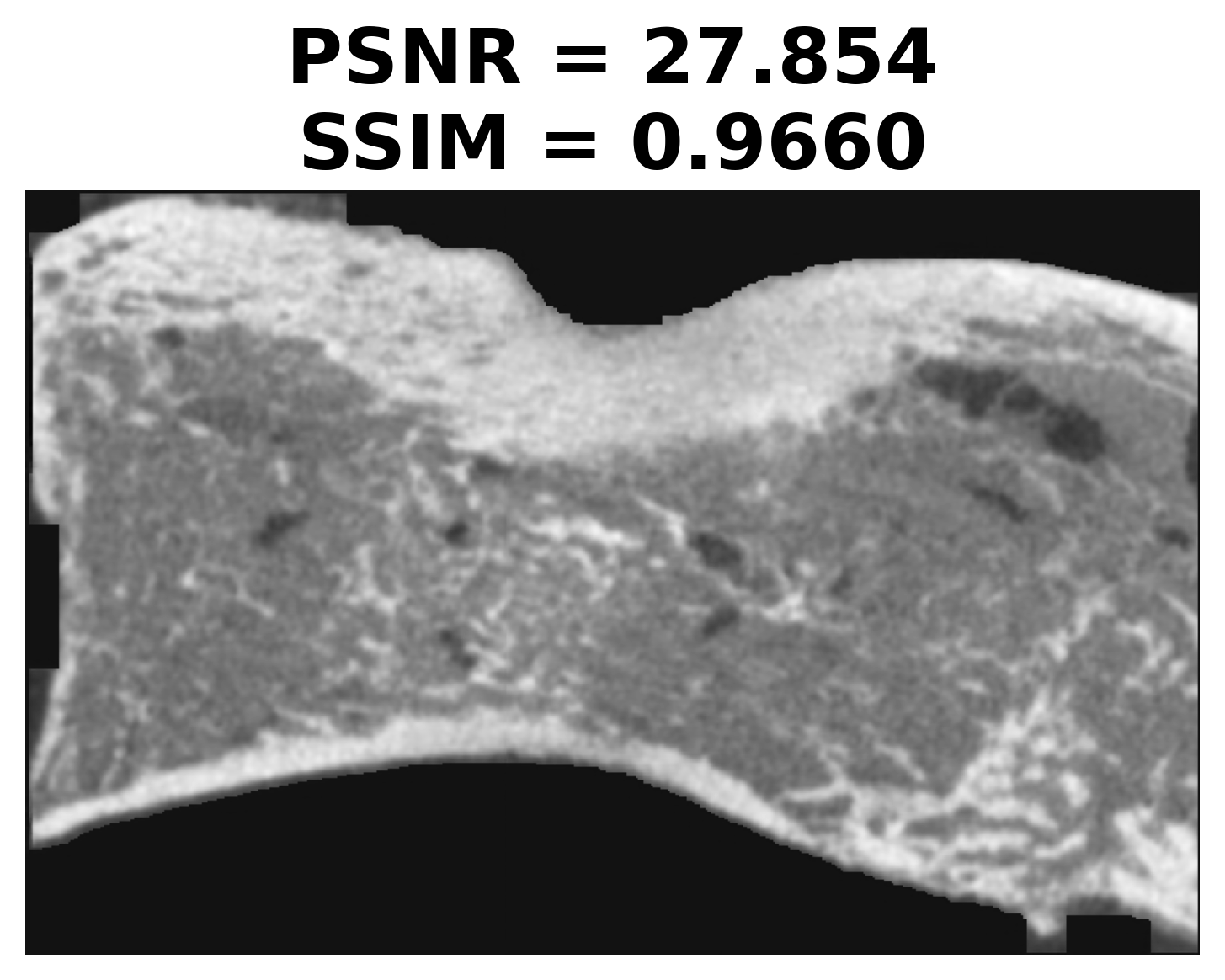}
		\end{subfigure}
		\begin{subfigure}[b]{0.16\textwidth}
			\includegraphics[width=\textwidth]{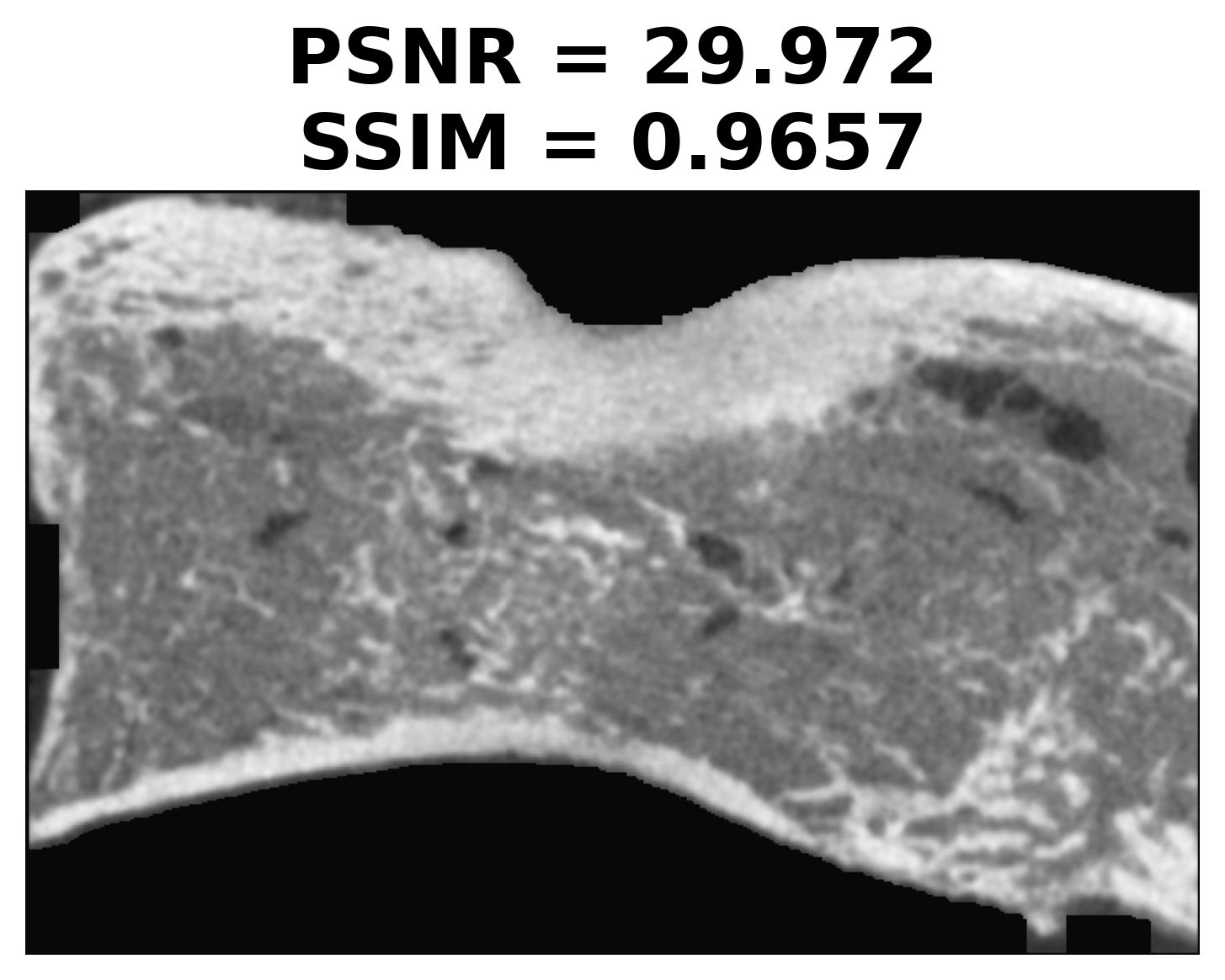}
		\end{subfigure}
		\begin{subfigure}[b]{0.16\textwidth}
			\includegraphics[width=\textwidth]{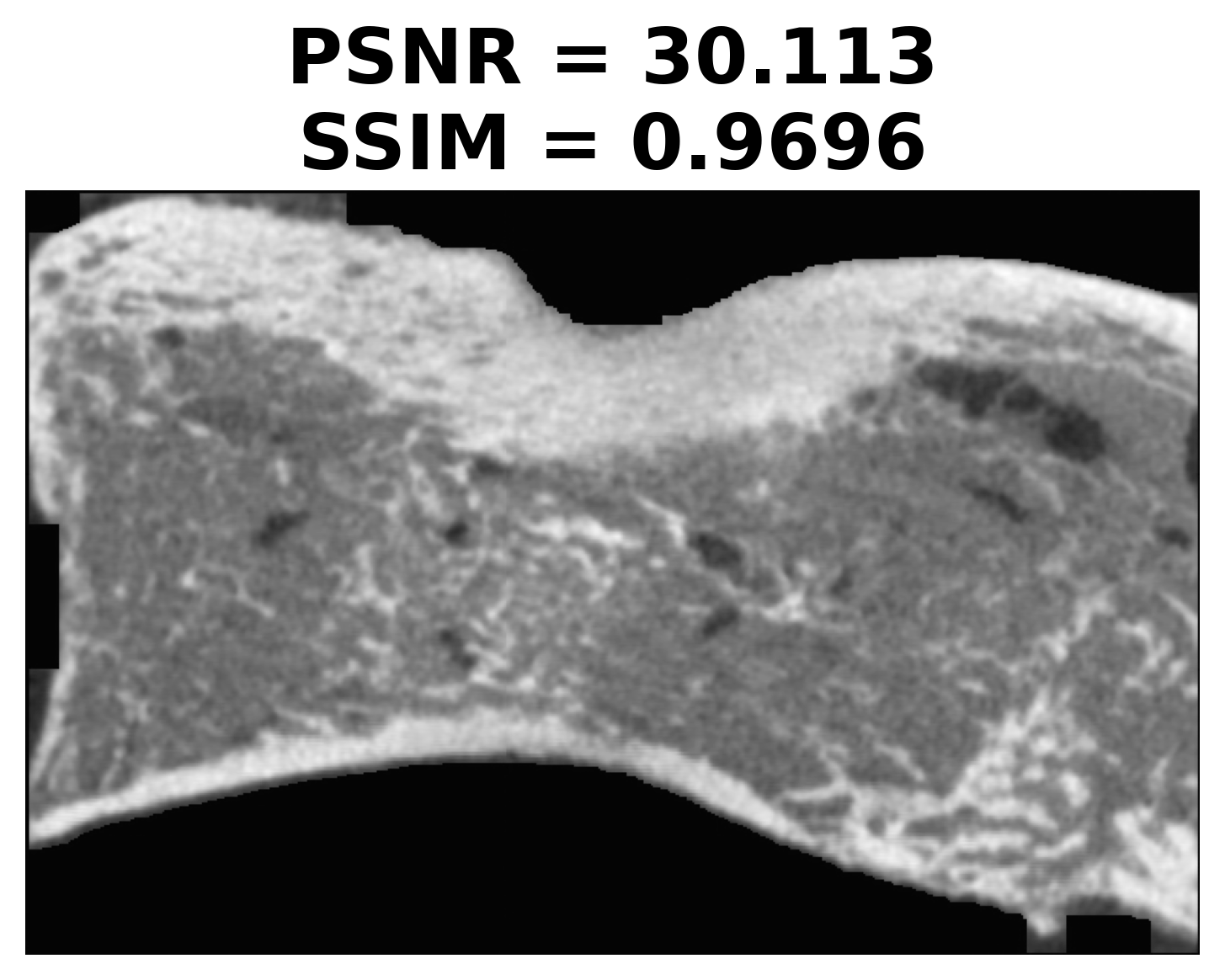}
		\end{subfigure}
		
		\begin{subfigure}[b]{0.16\textwidth}
			\includegraphics[width=\textwidth]{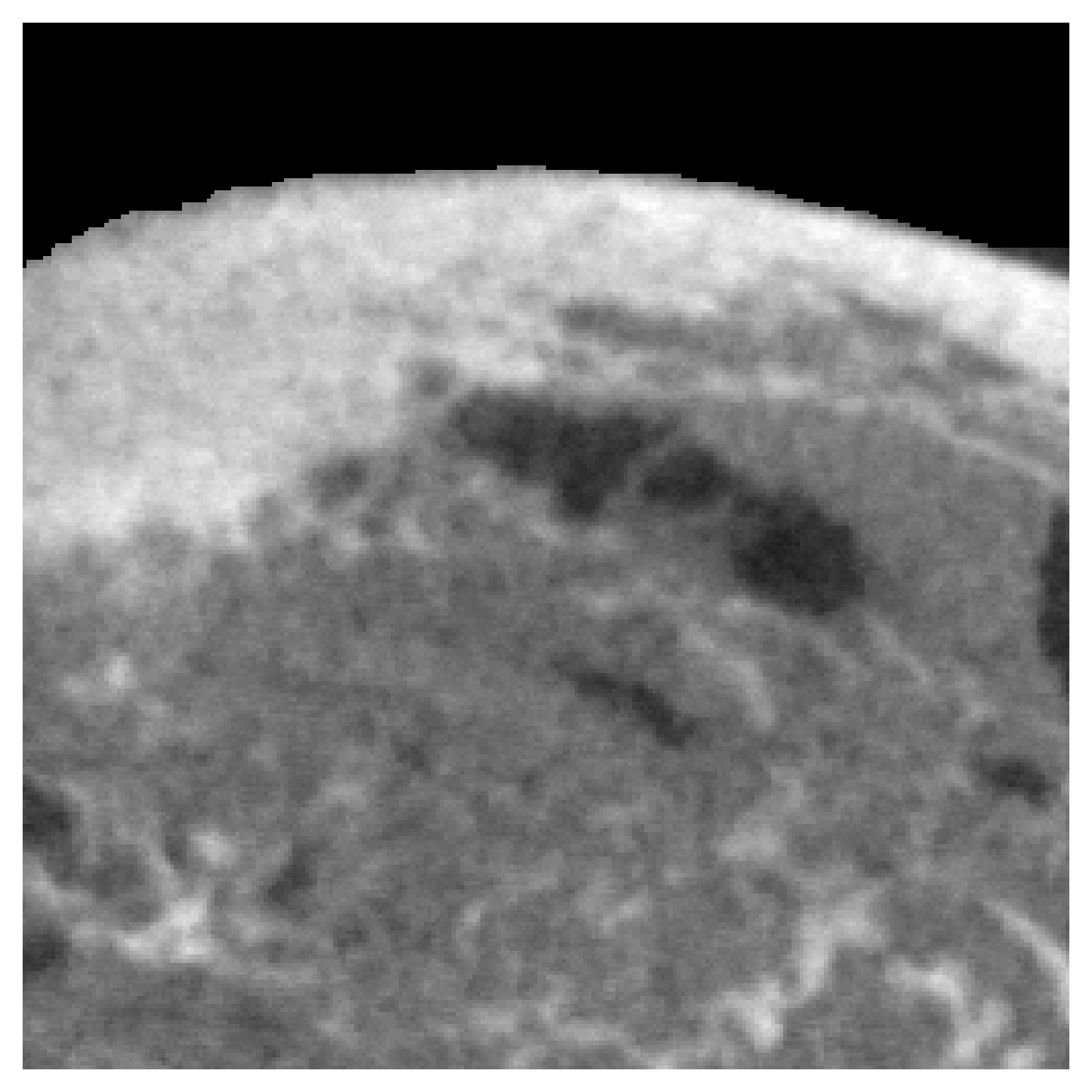}
			\caption{Noisy}
			\label{fig6a}
		\end{subfigure}
		\begin{subfigure}[b]{0.16\textwidth}
			\includegraphics[width=\textwidth]{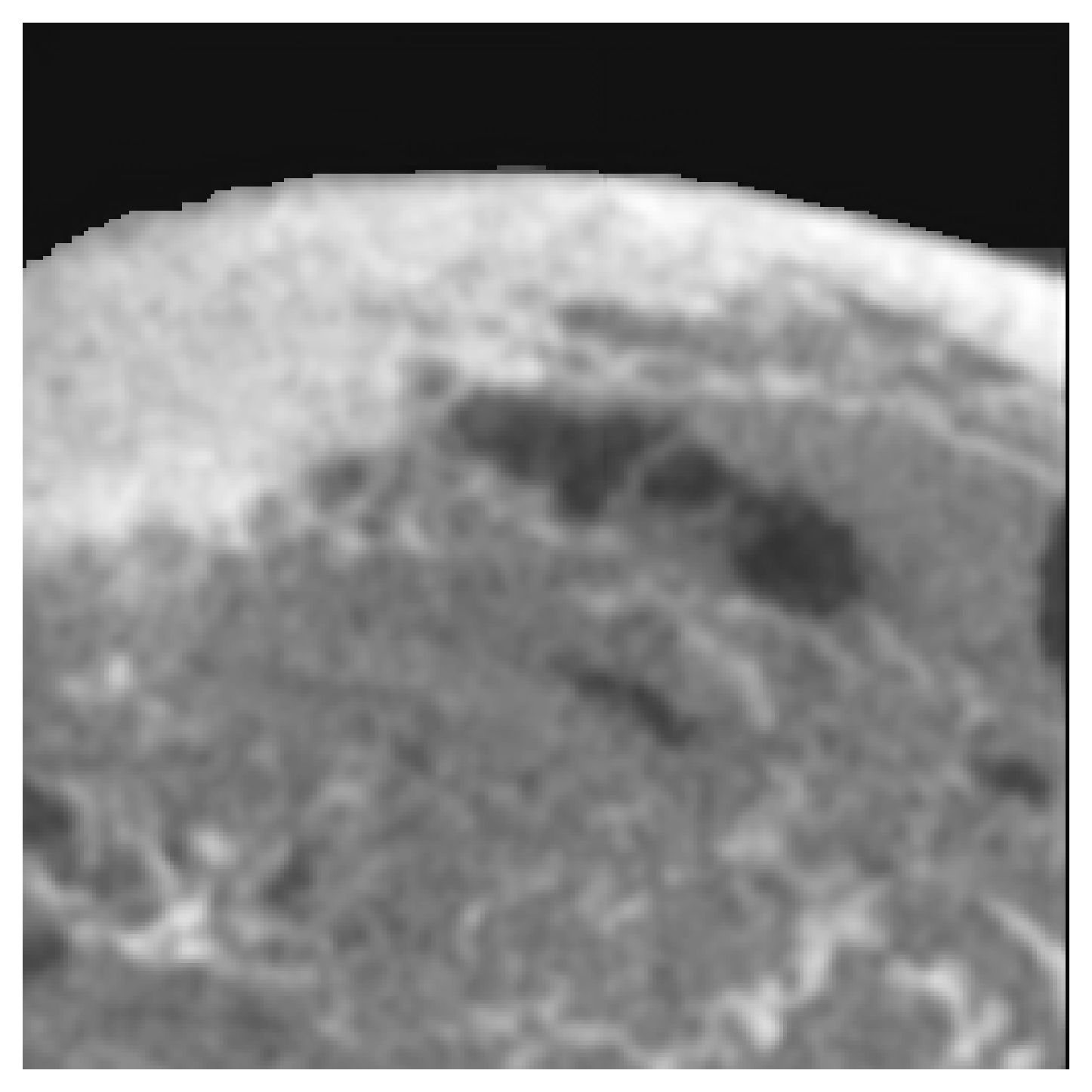}
			\caption{ResU-Net}
			\label{fig6b}
		\end{subfigure}
		\begin{subfigure}[b]{0.16\textwidth}
			\includegraphics[width=\textwidth]{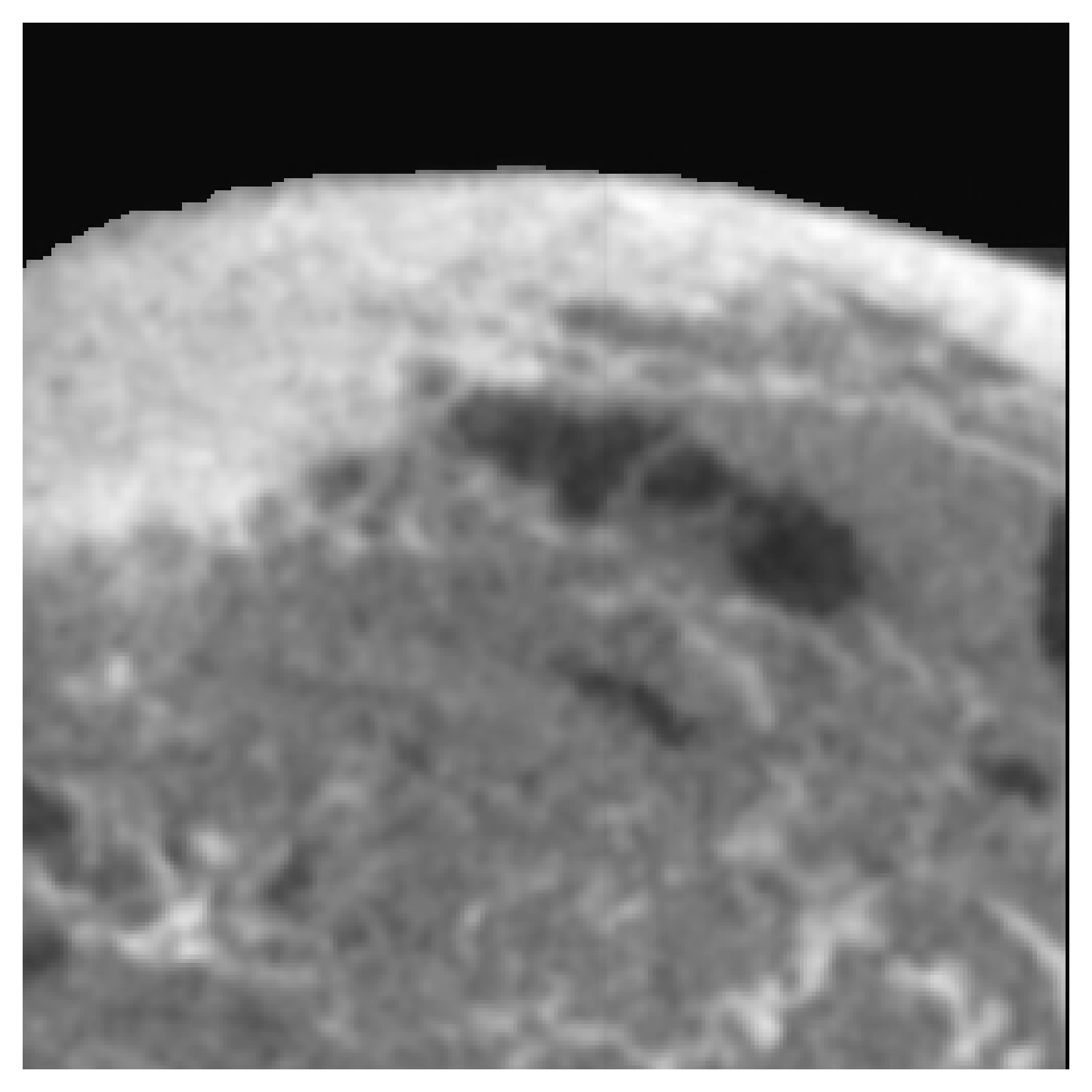}
			\caption{Uformer}
			\label{fig6c}
		\end{subfigure}
		\begin{subfigure}[b]{0.16\textwidth}
			\includegraphics[width=\textwidth]{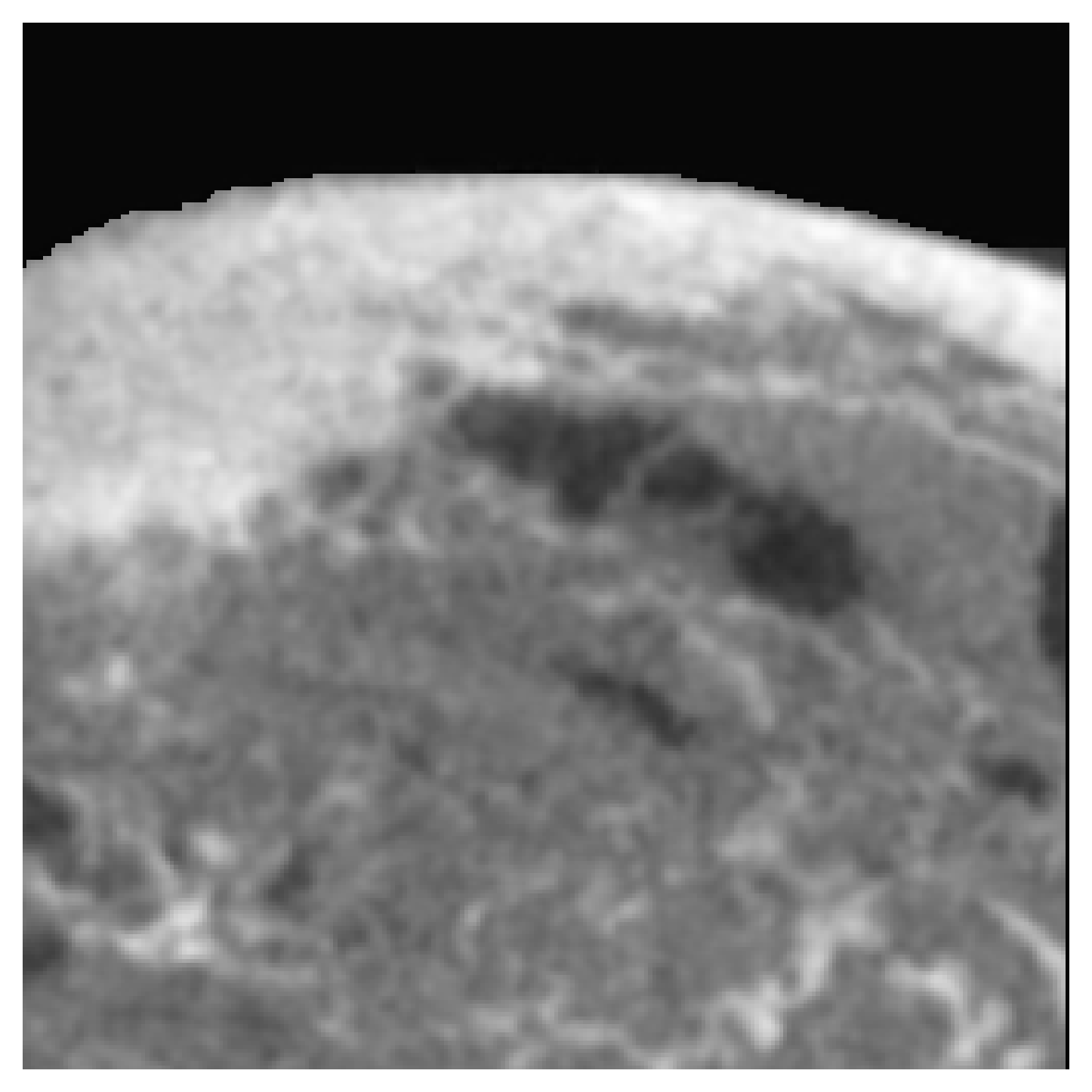}
			\caption{SwinIR}
			\label{fig6b}
		\end{subfigure}
		\begin{subfigure}[b]{0.16\textwidth}
			\includegraphics[width=\textwidth]{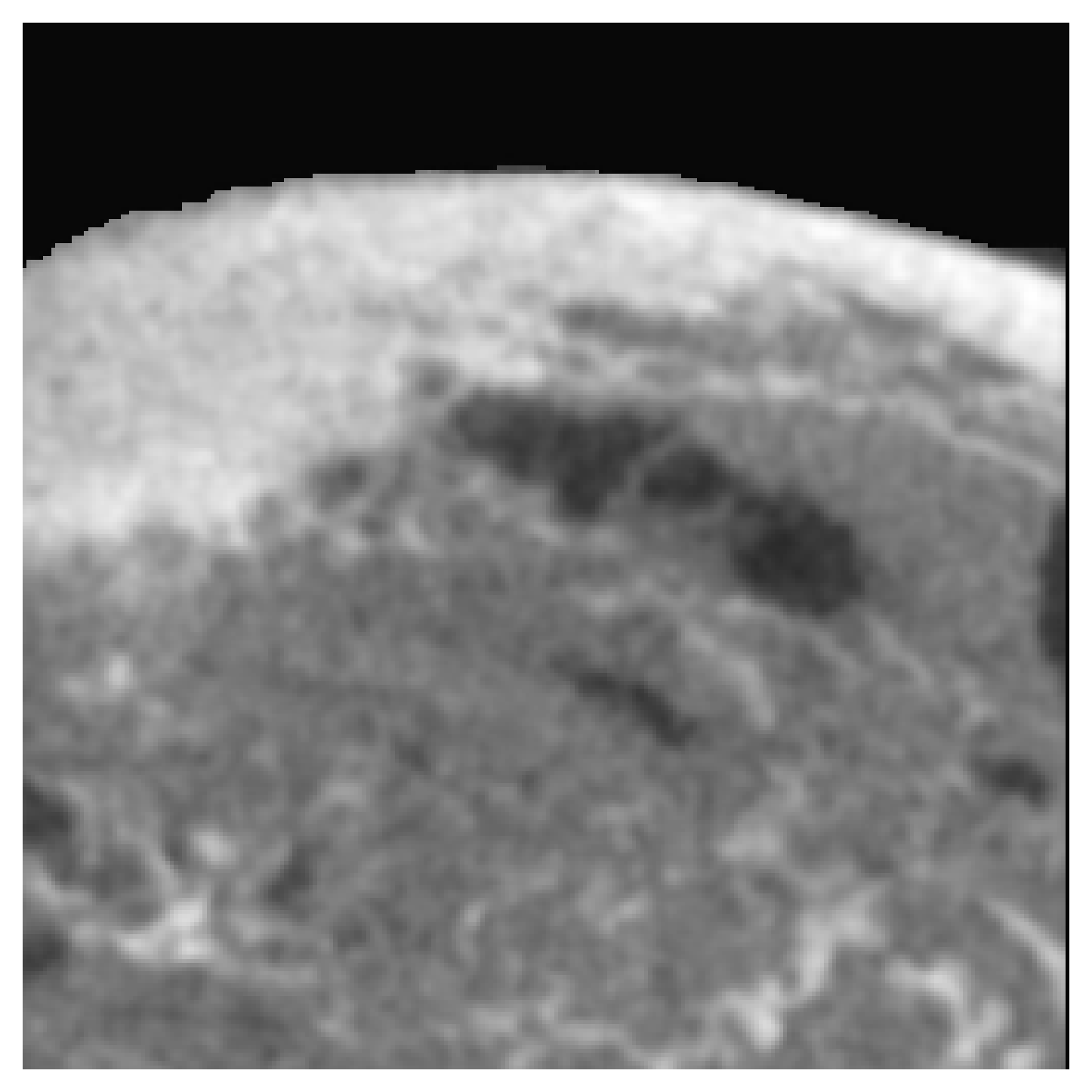}
			\caption{HAT}
			\label{fig6b}
		\end{subfigure}
		\begin{subfigure}[b]{0.16\textwidth}
			\includegraphics[width=\textwidth]{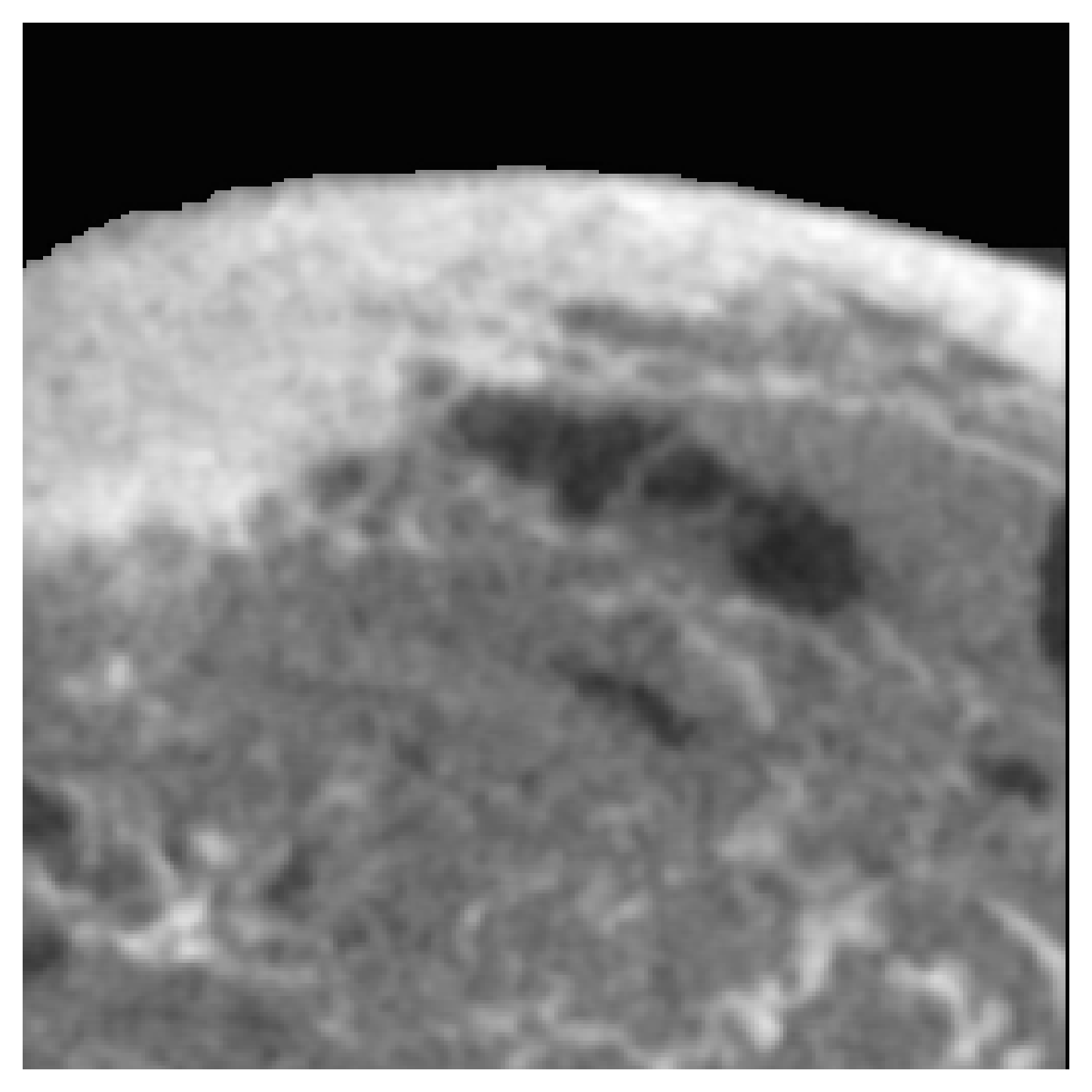}
			\caption{HARU-Net}
			\label{fig6b}
		\end{subfigure}
		\caption{Comparison of denoising performance on CBCT slice from frontal view from a 3D CBCT scan.}
		\label{fig06}
	\end{figure*}

	We next evaluate the qualitative performance across three representative CBCT slices from two different test samples, extracted from the sagittal, frontal, and axial anatomical views, as shown in Figures \ref{fig04}--\ref{fig06}. Figures \ref{fig04} and \ref{fig05} present the axial and sagittal views from the same scan, while Fig. \ref{fig06} shows the frontal view from another scan. We additionally provide zoomed-in views for each figure, focusing on relevant anatomical structures. Overall, ResU-Net, Uformer, SwinIR, and the proposed HARU-Net produce visually coherent and clinically interpretable depictions, effectively reducing noise while maintaining the visibility of key anatomical structures. However, HARU-Net produces the most visually consistent improvements, with clear enhancements in the sharpness of bone boundaries, cortical outlines, and internal trabecular patterns.
	
	In contrast, several of the competing approaches exhibit minor artifacts or structural inconsistencies despite their generally good performance. Although the visual improvements introduced by HARU-Net may appear modest at first glance, they are clearly observable in the zoomed regions and align well with the quantitative gains reported earlier.
	
	Overall, while all deep learning models provide clinically applicable reconstructions, HARU-Net demonstrates the most balanced and visually superior denoising performance, consistently delivering sharper and more anatomically faithful results across different slice orientations.
	
	\section{Discussion}
	Noise remains one of the major limiting factors in low-dose CBCT imaging, which serves as a primary modality used for dental, maxillofacial, and ENT diagnosis. Excessive noise often obscures fine anatomical structures such as root canal morphology and periapical lesions, potentially leading to diagnostic uncertainty or even misdiagnosis \cite{schulze2011artefacts,patel2019cone,rios2024cone}. Moreover, the clarity of bone structure may be reduced by noise, which adversely impacts the planning and placement of implants \cite{patel2019cone,pauwels2015effect,camilo2013artefacts}. In addition, cephalometric analysis and airway evaluation are susceptible to noise, making it difficult to measure anatomical landmarks precisely \cite{chung2024validation,lee2019effect}. The situation is further aggravated in dynamic CBCT applications, such as TMJ evaluations, where repeated scans are needed to capture joint motion \cite{bag2014imaging}. Additionally, tissue characterization through dual-energy CBCT suffers from metal artifacts and noise because of the complex decomposition process \cite{zhu2019evaluation}. 
	
	While DL has been widely applied for image enhancement and reconstruction in CT and MRI, its use for CBCT image enhancement remains limited, primarily due to the lack of paired low- and high-quality training data. This article addresses this gap by introducing a Hybrid Attention Residual U-Net (HARU-Net), trained using a cadaver-derived CBCT dataset acquired at higher radiation doses. Noisy and clean pairs were generated through a controlled noising process. 
	
	The experimental results demonstrate that the proposed HARU-Net substantially improves CBCT denoising performance compared with both the baseline U-Net architecture and state-of-the-art transformer-based models. This enhanced performance stems from the integration of transformer-style HABs at each skip connection and an RHAG at the bottleneck. These components combine the representational strengths of transformer blocks with the stability and computational efficiency of convolutional feature extraction within a unified architecture. This hybrid design proves particularly effective for CBCT denoising, where noise is spatially varying, signal-dependent, and structurally complex. While transformer-based models such as Uformer and SwinIR rely heavily on computationally intensive attention mechanisms to capture long-range dependencies, HARU-Net uses the strengths of a convolutional backbone to efficiently extract local features, while transformer-style attention blocks provide global context to augment the local representations. This results in superior restoration quality at substantially lower computational cost, highlighting the advantage of integrating selective transformer components within a CNN framework.
	
	The comparison with ResU-Net is particularly interesting because, while it exhibits the lowest computational cost due to its purely convolutional architecture, it also shows the worst denoising performance in terms of PSNR and GMSD. What makes this finding notable is that the incorporation of HABs into the skip connections and an RHAG at the bottleneck effectively elevates the capability of the baseline ResU-Net. These additions introduce only a modest increase in computational complexity yet yield a substantial improvement in denoising performance, demonstrating the effectiveness and efficiency of the proposed architectural enhancements. Regardless, our findings suggest that ResU-Net can serve as a baseline for developing future fast and effective real-time CBCT denoisers.
	
	We further demonstrate the clinical relevance of the proposed method through visual evaluation. Compared with transformer-dominant architectures, HARU-Net produces reconstructions that more faithfully preserve fine anatomical details without introducing over-smoothing or attention-induced artifacts. It must be stated that the denoised images from Uformer, SwinIR, and ResU-Net yield clinically relevant results, although HARU-Net appears visually superior. These improvements emphasize the value of integrating CNN-based inductive biases with transformer-style attention in a balanced and computationally efficient manner. These results suggest that such hybrid architectures offer a promising direction for developing high-performing yet lighter-weight denoising models for dentistry and broader medical imaging applications.
	
	Although computationally lighter than both Uformer and SwinIR, full-volume inference for a $512\times512\times512$ CBCT scan requires approximately 2 minutes on a consumer-grade GPU (NVIDIA RTX 2080 Ti). This is substantially faster than Uformer (approximately 4.30 minutes) and SwinIR (approximately 8.85 minutes), yet still falls short of real-time processing, which would facilitate clinical deployment. This limitation highlights the need for further exploration of model compression, architectural pruning, and efficient attention mechanisms to produce an even lighter variant of HARU-Net. Nevertheless, the present findings establish HARU-Net as an effective, computationally efficient, and clinically meaningful advancement toward high-fidelity denoising in low-dose CBCT imaging.
	
	A key limitation of the proposed approach is that the training data originate from a limited sample scanned on a device from a single CBCT vendor, which restricts the model’s generalizability across different scanner types. Future work should therefore assess cross-vendor performance and explore strategies such as vendor-specific fine-tuning or domain adaptation to strengthen the model’s robustness in clinical settings.
	
	\section{Conclusion}
	In this work, we have introduced a Hybrid Attention Residual U-Net tailored for denoising CBCT data. By integrating residual convolutional encoding with hybrid attention mechanisms both within the bottleneck and along skip connections, HARU-Net effectively captures both local anatomical detail and global contextual structure. Comprehensive experiments using real CBCT scans demonstrate that HARU-Net consistently outperforms several state-of-the-art denoising models, including SwinIR, Uformer, and residual U-Net baselines, achieving higher PSNR, SSIM, and GMSD scores, as well as superior perceptual quality across multiple anatomical views. Overall, HARU-Net offers a fast, accurate, and clinically viable solution for enhancing CBCT image quality. Its ability to preserve fine anatomical structures while effectively suppressing noise positions it as a promising candidate for real-time CBCT enhancement workflows in dental, maxillofacial, and ENT imaging. Future work will explore adapting the model for volumetric (3D) denoising, multi-site generalization, and integration with self-supervised or physics-informed learning frameworks.
	
	
	\small
	\bibliographystyle{flairs}
	\bibliography{CBCT_Denoising_References}

\end{document}